\documentclass[journal]{IEEEtran}

\usepackage{graphicx,latexsym,color}
\usepackage{amsmath,amsthm,amsfonts}
\usepackage{latexsym}
\usepackage{cite,citesort}

\usepackage{bm}

\renewcommand{\vec}[1]{{\boldsymbol#1}}

\newcommand{\ie}{\textit{i.e.}\/, }
\newcommand{\eg}{\textit{e.g.}\/, }

\providecommand*{\mrm}[1]{\mathrm{#1}}
\providecommand*{\unit}[1]{\ensuremath{\mrm{\,#1}}}
\providecommand*{\eu}{\ensuremath{\mrm{e}}}
\providecommand*{\iu}{\ensuremath{\mrm{i}}}

\providecommand*{\diff}{\operatorname{d}\!}
\renewcommand{\Re}{\operatorname{Re}}	
\renewcommand{\Im}{\operatorname{Im}}	
\providecommand*{\ohm}{\ensuremath{\mrm{\Omega}}}

\providecommand*{\degree}{\ensuremath{^\circ}}

\begin{document}

\title{Dispersion modeling and analysis for multilayered open coaxial waveguides}

\author{Sven~Nordebo,~\IEEEmembership{Senior Member,~IEEE}, G\"{o}khan~Cinar, Stefan~Gustafsson, B\"{o}rje~Nilsson
\thanks{Manuscript received \today. This work was supported in part by the Swedish Research Council (VR) and ABB AB.}  
\thanks{Sven~Nordebo and Stefan~Gustafsson are with the Department of Physics and Electrical Engineering, Linn\ae us University, 
351 95 V\"{a}xj\"{o}, Sweden. (E-mail: \{sven.nordebo,stefan.h.gustafsson\}@lnu.se).}
\thanks{G\"{o}khan~Cinar is with the Electronics Engineering Department, Gebze Institute of Technology, 
414 00 Gebze, Kocaeli, Turkey. (E-mail: gcinar@gmail.com).}
\thanks{B\"{o}rje~Nilsson is with the Department of Mathematics, Linn\ae us University, 
351 95 V\"{a}xj\"{o}, Sweden. (E-mail: borje.nilsson@lnu.se).}
}


\maketitle

\begin{abstract}
This paper presents a detailed modeling and analysis regarding the 
dispersion characteristics of multilayered open coaxial waveguides. The study is motivated
by the need of improved modeling and an increased physical understanding about the wave propagation phenomena
on very long power cables which has a potential industrial application with fault localization and monitoring.
The electromagnetic model is based on a layer recursive computation of axial-symmetric fields in connection
with a magnetic frill generator excitation that can be calibrated to the current measured at the input of the cable.
The layer recursive formulation enables a stable and efficient numerical computation of the related dispersion functions as well as
a detailed analysis regarding the analytic and asymptotic properties of the associated determinants.
Modal contributions as well as the contribution from the associated branch-cut (non-discrete radiating modes) 
are defined and analyzed.
Measurements and modeling of pulse propagation on
an 82 km long HVDC power cable are presented as a concrete example.
In this example, it is concluded that the contribution from the second TM mode as well as from 
the branch-cut is negligible for all practical purposes. However, it is also shown that for extremely long power cables the
contribution from the branch-cut can in fact dominate over the quasi-TEM mode for some frequency intervals.
The main contribution of this paper is to provide the necessary analysis tools for a quantitative study of these phenomena.
\end{abstract}

\begin{IEEEkeywords}
Submarine power cables, guided waves, transmission lines, dispersion relations, open waveguides.
\end{IEEEkeywords}

\section{Introduction}\label{sect:introduction}

\IEEEPARstart{T}{}he
topic of this paper is to provide a detailed modeling and analysis regarding the 
dispersion characteristics of multilayered open coaxial waveguides. This topic is motivated
in particular by the need of improved modeling and an increased physical understanding about the wave propagation phenomena
on very long power cables which has a potential industrial application with fault localization and monitoring.
Application examples include transient signal analysis and partial discharge diagnostics to enable cable maintenance and repair without
(or with very short) power losses, see \eg \cite{Boggs+etal1996,Pommerenke+etal1999,Stone2005,Veen2005}.
Hence, there is today an increasing interest to improve both measurement techniques and modeling regarding
the wave propagation characteristics of power cables and its dependency on various material and structural 
parameters, see \eg \cite{Gustavsen+etal1999,Heinrich+etal2000,Amekawa+etal2003,Steinbrich2005,Gustavsen+etal2005,Papazyan+etal2007,Yang+Zhang2007,Gudmundsdottir+etal2011,Nordebo+etal2013a,Nordebo+etal2013b}.

It is also the aim of this paper to address some basic questions regarding the analytical properties
of modes and branch-cuts (non-discrete radiating modes). 
Many interesting general properties of modes and branch-cuts can be considered from a purely theoretical point of view
see \eg \cite{Collin1991,Felsen+Marcuvitz1994,Olyslager1999,Nosich1994,Nosich2000}.
In particular, it has been shown in \cite{Nosich1994,Nosich2000} that for an open waveguide structure and under very general conditions, modes do exist
as poles of a meromorphic Fourier transform, the poles have only finite multiplicity, they have no finite accumulation point, they are
continuous functions of geometry, material parameters and frequency (unless two or more poles coalesce), and finally that poles
can appear and disappear only at the boundary of the domain of meromorphicity, \ie at infinity and in the branch-point corresponding
to the wavenumber of the exterior domain. When there are sources present, 
the solution is obtained by taking the Fourier transform in the longitudinal direction of the waveguide, followed by residue calculus. 
The discrete set of eigenfunctions are obtained as residues of poles and the non-discrete set is manifested as an integration 
at the branch-cut \cite{Collin1991,Felsen+Marcuvitz1994,Olyslager1999}.
It is the aim of the present paper to provide the necessary analytical and numerical tools for a quantitative study of these phenomena
in the important special case with a multilayered circular geometry including non-perfect conductors (metallic layers) at low frequencies.

The electromagnetic model here is based on a layer recursive computation of axial-symmetric fields in connection with 
a magnetic frill generator excitation that can be calibrated to the current measured at the input of the cable. 
Inverse Fourier transforms are executed in the spatial domain by using numerical contour integration and in the time domain by using an 
inverse FFT (Fast Fourier Transform). 
The electromagnetic model is validated by comparing with measurements of pulse propagation on an 82 km long HVDC power cable 
in the low frequency range of 0-100 kHz, see also \cite{TEAT-7211}. 
Previously, the layer recursive formulation has been used to study the low frequency behavior of the dominating pole
of a multilayered coaxial cable \cite{Nordebo+etal2013b}, and asymptotic theory has been used to study the contribution from
the branch-cut regarding the spectral behavior of a single core wire in an open domain \cite{Nordebo+etal2013a}.
In the present contribution, several new modeling aspects are discussed in detail such as the recursive computation of determinants
in connection with an exponential scaling of Bessel functions to achieve a stable and efficient computation of fields inside the metals,
as well as the magnetic frill generator model which is also commonly used as an excitation model for antennas \cite{Volakis+Sertel2011}. 
Furthermore, the layer recursive formulation in connection with the magnetic frill excitation enables a detailed analysis regarding 
the analytic and asymptotic properties of the currents inside the waveguide (in terms of the complex valued longitudinal
Fourier variable), at infinity as well as at the branch-point. The application of Jordan's lemma is hence justified,
and the contributions from the poles as well as from the branch-cut is established, and can be computed numerically. 
The analysis finally enables a derivation of the asymptotic properties of the integral at the branch-cut.

\section{Electromagnetic modeling of axial-symmetric fields}
\subsection{Fields, material parameters and geometry}
Let $\mu_0$, $\epsilon_0$, $\eta_0$ and ${\rm c}_0$ denote the permeability, the permittivity, the wave impedance and
the speed of light in vacuum, respectively, and where $\eta_0=\sqrt{\mu_0/\epsilon_0}$ and ${\rm c}_0=1/\sqrt{\mu_0\epsilon_0}$.
The wavenumber of vacuum is given by $k_0=\omega/{\rm c}_0$ where $\omega=2\pi f$ is the angular frequency and $f$ the frequency.
It is also convenient to use $\omega\mu_0=k_0\eta_0$ and $\omega\epsilon_0=k_0/\eta_0$.
The cylindrical coordinates are denoted by $(\rho,\phi,z)$, the corresponding unit vectors $(\hat{\vec{\rho}},\hat{\vec{\phi}},\hat{\vec{z}})$,
the transverse coordinate vector $\vec{\rho}=\rho\hat{\vec{\rho}}$ and the radius vector $\vec{r}=\vec{\rho}+z\hat{\vec{z}}$.  

Let $\vec{E}(\vec{r})$ and $\vec{H}(\vec{r})$ denote the electric and magnetic fields, respectively, 
and where the time-harmonic factor  $\eu^{-\iu\omega t}$ has been suppressed. The corresponding
Maxwell's equations \cite{Jackson1999} for the fields inside a general isotropic material are given by
\begin{equation}\label{eq:Maxwell1}
\left\{\begin{array}{l}
\nabla\times \vec{E}(\vec{r})=-\vec{M}(\vec{r})+\iu\omega\mu_0\mu\displaystyle\vec{H}(\vec{r}), \vspace{0.2cm}\\
\nabla\times\vec{H}(\vec{r})=-\iu\omega\epsilon_0\epsilon\vec{E}(\vec{r}),
\end{array}\right.
\end{equation}
where $\mu$ and $\epsilon$ are the complex valued relative permeability and permittivity of the material, respectively,
and $\vec{M}(\vec{r})$ the magnetic current density of the impressed source (excitation).

Following the definition of cylindrical vector waves as defined in Appendix \ref{app:cyldef}, the axial-symmetric ($m=0$)
transverse magnetic (TM) fields in a homogeneous and isotropic cylindrical region can be expressed as
\begin{eqnarray}
E_{\rho}(\rho,\alpha)=-\frac{\iu\alpha}{k_0}\left[ a {\rm J}_1(\kappa\rho)+b {\rm H}_1^{(1)}(\kappa\rho)\right],  \label{eq:Er} \\
E_{z}(\rho,\alpha)=\frac{1}{k_0}\left[ a \kappa {\rm J}_0(\kappa\rho)+b\kappa  {\rm H}_0^{(1)}(\kappa\rho)\right],  \label{eq:Ez} \\
H_{\phi}(\rho,\alpha)=\frac{1}{\iu\eta_0}\left[ a \epsilon {\rm J}_1(\kappa\rho)+b \epsilon {\rm H}_1^{(1)}(\kappa\rho)\right],  \label{eq:Hphi}
\end{eqnarray}
where $\alpha$ is the complex valued Fourier variable corresponding to the propagation factor $\eu^{\iu\alpha z}$, and
${\rm J}_m(\kappa\rho)$ and ${\rm H}_m^{(1)}(\kappa\rho)$ are the regular Bessel functions and 
the Hankel functions of the first kind, both of order $m$, respectively. Here, the
transverse wavenumber is defined by
\begin{equation}\label{eq:kappadef}
\kappa=\sqrt{k_0^2\mu\epsilon-\alpha^2},
\end{equation}
where the square root\footnote{If the square root is defined as \eg with the MATLAB software where
$-\pi/2<\arg\sqrt{w}\leq\pi/2$ for $-\pi<\arg w\leq \pi$, then $\kappa$ can be defined here as $\kappa=\iu\sqrt{-k_0^2\mu\epsilon+\alpha^2}$
which implies that $0<\arg\kappa\leq \pi$.} $\kappa=\sqrt{w}$ is defined such that $0<\arg w\leq 2\pi$ and
$0<\arg\kappa\leq \pi$ and hence $\Im\kappa\geq 0$. Furthermore, the expansion coefficients have been chosen as
$a=a_{20}/\sqrt{\mu\epsilon}$ and $b=b_{20}/\sqrt{\mu\epsilon}$ for notational convenience (see Appendix \ref{app:cyldef}), and the relationship
${\rm C}_{0}^\prime(\zeta)=-{\rm C}_{1}(\zeta)$ have been used which is valid for all Bessel functions ${\rm C}_m(\zeta)$ \cite{Olver+etal2010}.

Consider now a multilayered open coaxial waveguide with $N$ cylindrical layers with radius $\rho_i$,
and $N+1$ material regions as depicted in Fig.~\ref{fig:matfigslayers} below.
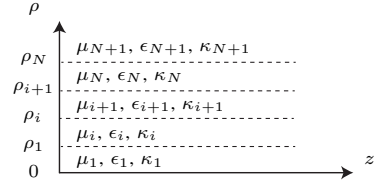
\begin{figure}[htb]
\begin{picture}(50,70)
\put(100,0){\makebox(50,60){\includegraphics[width=5cm]{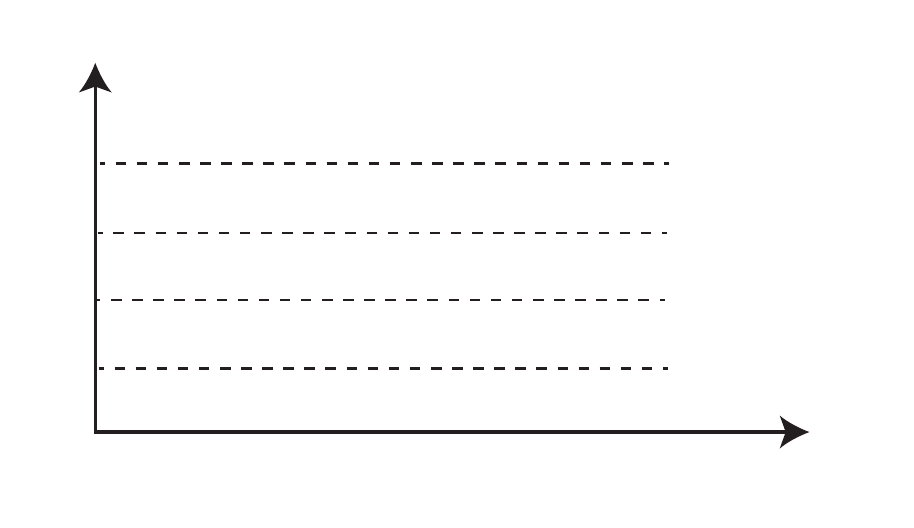}}} 
\put(57,65){\scriptsize $\rho$}
\put(54,46){\scriptsize $\rho_{N}$}
\put(51,35){\scriptsize $\rho_{i+1}$}
\put(55,24){\scriptsize $\rho_{i}$}
\put(55,13){\scriptsize $\rho_1$}
\put(57,2){\scriptsize $0$}
\put(184,7){\scriptsize $z$}
\put(75,50){\scriptsize $\mu_{N+1}$, $\epsilon_{N+1}$, $\kappa_{N+1}$}
\put(75,39){\scriptsize $\mu_N$, $\epsilon_N$, $\kappa_N$}
\put(75,28){\scriptsize $\mu_{i+1}$, $\epsilon_{i+1}$, $\kappa_{i+1}$}
\put(75,17){\scriptsize $\mu_i$, $\epsilon_i$, $\kappa_i$}
\put(75,7){\scriptsize $\mu_1$, $\epsilon_1$, $\kappa_1$}
\end{picture}
\caption{Definition of material regions.}
\label{fig:matfigslayers}
\end{figure}

The material parameters $\epsilon_i$, $\mu_i$ and the corresponding transverse wavenumbers $\kappa_i=\sqrt{k_0^2\mu_i\epsilon_i-\alpha^2}$ 
are assumed to be fixed for each cylindrical region and numbered by $i=1,\ldots,N+1$. The interior region is defined by $\rho<\rho_1$,
the intermediate regions by $\rho_{i-1}<\rho<\rho_i$ for $i=2,\ldots,N$, and the exterior region by $\rho>\rho_N$. 
In this paper\footnote{All theoretical results and numerical implementations in this paper can straightforwardly be modified for any other
frequency dependent material dispersion model such as Debye, Lorentz, Drude, etc.}, the complex valued relative permittivity of each layer is given by
\begin{equation}
\epsilon_i=\epsilon_{{\rm r}i}+\iu\frac{\sigma_i}{\omega\epsilon_0},
\end{equation}
where $\epsilon_{{\rm r}i}$ is the corresponding real relative permittivity and  $\sigma_i$ the conductivity of the material.
The relative permeability $\mu_i$ is assumed to be real valued and frequency independent.

\subsection{Excitation model and explicit solutions}
As an excitation model is used a
magnetic frill generator with magnetic surface current $\vec{M}_{\rm S}=M(\omega)\delta(z)\hat{\vec{\phi}}$
at the inner conductor boundary at $\rho_1$. Here $M(\omega)$ is the excitation amplitude in the frequency domain and 
$\delta(z)$ denotes the spatial Dirac delta distribution.
Let $a_1$, $a_i$ and $b_i$ for $i=2,\ldots,N$ and $b_{N+1}$ denote the $2N$ expansion coefficients corresponding to the $N+1$ material regions
defined as in \eqref{eq:Er} through \eqref{eq:Hphi} above. 
The boundary conditions related to the inner boundary at radius $\rho_1$ are then given by employing the continuity (and discontinuity) properties
of the tangential fields $(E_z,H_{\phi})$ in \eqref{eq:Ez} and \eqref{eq:Hphi} to yield
\begin{equation}\label{eq:Matconds1}
\left\{\begin{array}{l}
  -a_1\kappa_{1}{\rm J}_0(\kappa_1\rho_1) \vspace{0.2cm}\\
 +a_2\kappa_{2}{\rm J}_0(\kappa_{2}\rho_1)
 +b_2\kappa_{2}{\rm H}_0^{(1)}(\kappa_{2}\rho_1)
= k_0M(\omega), \vspace{0.2cm}\\
 -a_1\epsilon_{1} {\rm J}_1(\kappa_1\rho_1) \vspace{0.2cm}\\
  +a_2\epsilon_{2} {\rm J}_1(\kappa_{2}\rho_1)
+b_2\epsilon_{2}{\rm H}_1^{(1)}(\kappa_{2}\rho_1)
= 0.
\end{array}\right.
\end{equation}
The boundary conditions related to the intermediate boundaries at radius $\rho_{i}$ are similarly given by
\begin{equation}\label{eq:Matconds2}
\left\{\begin{array}{l}
  -a_{i}\kappa_{i}{\rm J}_0(\kappa_{i}\rho_{i})-b_{i}\kappa_{i}{\rm H}_0^{(1)}(\kappa_{i}\rho_{i}) \vspace{0.2cm}\\
 +a_{i+1}\kappa_{i+1}{\rm J}_0(\kappa_{i+1}\rho_{i})+b_{i+1}\kappa_{i+1}{\rm H}_0^{(1)}(\kappa_{i+1}\rho_{i})
= 0, \vspace{0.2cm}\\
-a_{i}\epsilon_{i}{\rm J}_1(\kappa_{i}\rho_{i})-b_{i}\epsilon_{i}{\rm H}_1^{(1)}(\kappa_{i}\rho_{i}) \vspace{0.2cm}\\
+a_{i+1}\epsilon_{i+1}{\rm J}_1(\kappa_{i+1}\rho_{i})+b_{i+1}\epsilon_{i+1}{\rm H}_1^{(1)}(\kappa_{i+1}\rho_{i})
= 0,
\end{array}\right.
\end{equation}
where $i=2,\ldots,N-1$. The boundary conditions related to the outer boundary at radius $\rho_{N}$ are finally given by
\begin{equation}\label{eq:Matconds3}
\left\{\begin{array}{l}
-a_{N}\kappa_{N}{\rm J}_0(\kappa_{N}\rho_{N})-b_{N}\kappa_{N}{\rm H}_0^{(1)}(\kappa_{N}\rho_{N}) \vspace{0.2cm}\\
+ b_{N+1}\kappa_{N+1}{\rm H}_0^{(1)}(\kappa_{N+1}\rho_{N})
= 0, \vspace{0.2cm}\\
-a_{N}\epsilon_{N}{\rm J}_1(\kappa_{N}\rho_{N})-b_{N}\epsilon_{N}{\rm H}_1^{(1)}(\kappa_{N}\rho_{N}) \vspace{0.2cm}\\
+ b_{N+1}\epsilon_{N+1}{\rm H}_1^{(1)}(\kappa_{N+1}\rho_{N})
= 0.
\end{array}\right.
\end{equation}

By using the Cramer's rule\cite{Strang1988}, the solution to the linear system of equations in \eqref{eq:Matconds1} through \eqref{eq:Matconds3}
can be written as
\begin{equation}\label{eq:a1sol}
a_1=k_0M(\omega)\frac{\det {\bf B}(\omega,\alpha)}{\det {\bf A}(\omega,\alpha)},
\end{equation}
where ${\bf A}(\omega,\alpha)$ and ${\bf B}(\omega,\alpha)$ are the corresponding square system matrices and where the first column of 
${\bf B}(\omega,\alpha)$ has been replaced with the unit vector $(1,0,\ldots,0)$.
The dispersion equation for root finding is given by $\det {\bf A}(\omega,\alpha)=0$, and
$\{\alpha_p\}_{p=1}^{\infty}$ will denote the corresponding set of poles. 

The Fourier representation of the current in the inner conductor is given by
\begin{multline}\label{eq:Iomegaalphadef}
I(\omega,\alpha)=\int_{0}^{2\pi}\int_{0}^{\rho_1}\sigma_1E_z(\rho,\alpha)\rho\diff\rho\diff\phi \\
=2\pi\sigma_1\frac{1}{k_0}a_1\kappa_1\int_{0}^{\rho_1}{\rm J}_0(\kappa_1\rho)\rho\diff\rho \\
=2\pi\sigma_1\frac{1}{k_0}a_1\rho_1{\rm J}_1(\kappa_1\rho_1),
\end{multline}
where $E_z(\rho,\alpha)$ is given by (\ref{eq:Ez}) and
where the integral $\int C_0(\zeta)\zeta\diff \zeta=\zeta C_1(\zeta)$ has been used \cite{Olver+etal2010}.
The current in the inner conductor at length position $z$ is hence given by
\begin{multline}\label{eq:Iomegazdef}
I(\omega,z)=\frac{1}{2\pi}\int_{-\infty}^{\infty}I(\omega,\alpha)\eu^{\iu\alpha z}\diff\alpha\\
=M(\omega)\sigma_1\rho_1\int_{-\infty}^{\infty}\frac{\det {\bf B}(\omega,\alpha)}{\det {\bf A}(\omega,\alpha)}
{\rm J}_1(\kappa_1\rho_1)\eu^{\iu\alpha z}\diff\alpha.
\end{multline}
The magnetic frill generator excitation amplitude $M(\omega)$ can now be calibrated from knowledge about (or measurements of) the
input current at the waveguide at $z=0+$, \ie $I_{\rm in}(\omega)=I(\omega,0+)$. Hence
\begin{equation}\label{eq:Mdef}
M(\omega)=\frac{I_{\rm in}(\omega)}{\sigma_1\rho_1\int_{-\infty}^{\infty}\frac{\det {\bf B}(\omega,\alpha)}{\det {\bf A}(\omega,\alpha)}
{\rm J}_1(\kappa_1\rho_1)\diff\alpha}.
\end{equation}
The current $I(\omega,z)$ can now be calculated at an arbitrary position $z$ by using  \eqref{eq:Iomegazdef}, or
\begin{equation}\label{eq:intFdef}
I(\omega,z)=\int_{-\infty}^{\infty}F(\alpha)\eu^{\iu\alpha z}\diff\alpha,
\end{equation}
where
\begin{equation}\label{eq:Fdef}
F(\alpha)=M(\omega)\sigma_1\rho_1\frac{\det {\bf B}(\omega,\alpha)}{\det {\bf A}(\omega,\alpha)}{\rm J}_1(\kappa_1\rho_1).
\end{equation}
Note that a simplified notation is used here where $F(\alpha)=\frac{1}{2\pi}I(\omega,\alpha)$
depends also on the frequency $\omega$.

The function $F(\alpha)$ is meromorphic with a sequence of poles $\{\alpha_p\}_{p=1}^{\infty}$, 
and the domain of meromorphicity is defined by a branch-cut at $\alpha_{\rm c}=k_0\sqrt{\mu_{N+1}\epsilon_{N+1}}$ 
corresponding to the wavenumber of the exterior region, see also \cite{Nosich1994,Nosich2000}.
As will be shown in section \ref{sect:analytprop}, $\alpha_{\rm c}$ is the only branch point of $F(\alpha)$.

\begin{figure}[htb]
\begin{picture}(50,120)
\put(100,0){\makebox(50,100){\includegraphics[width=8cm]{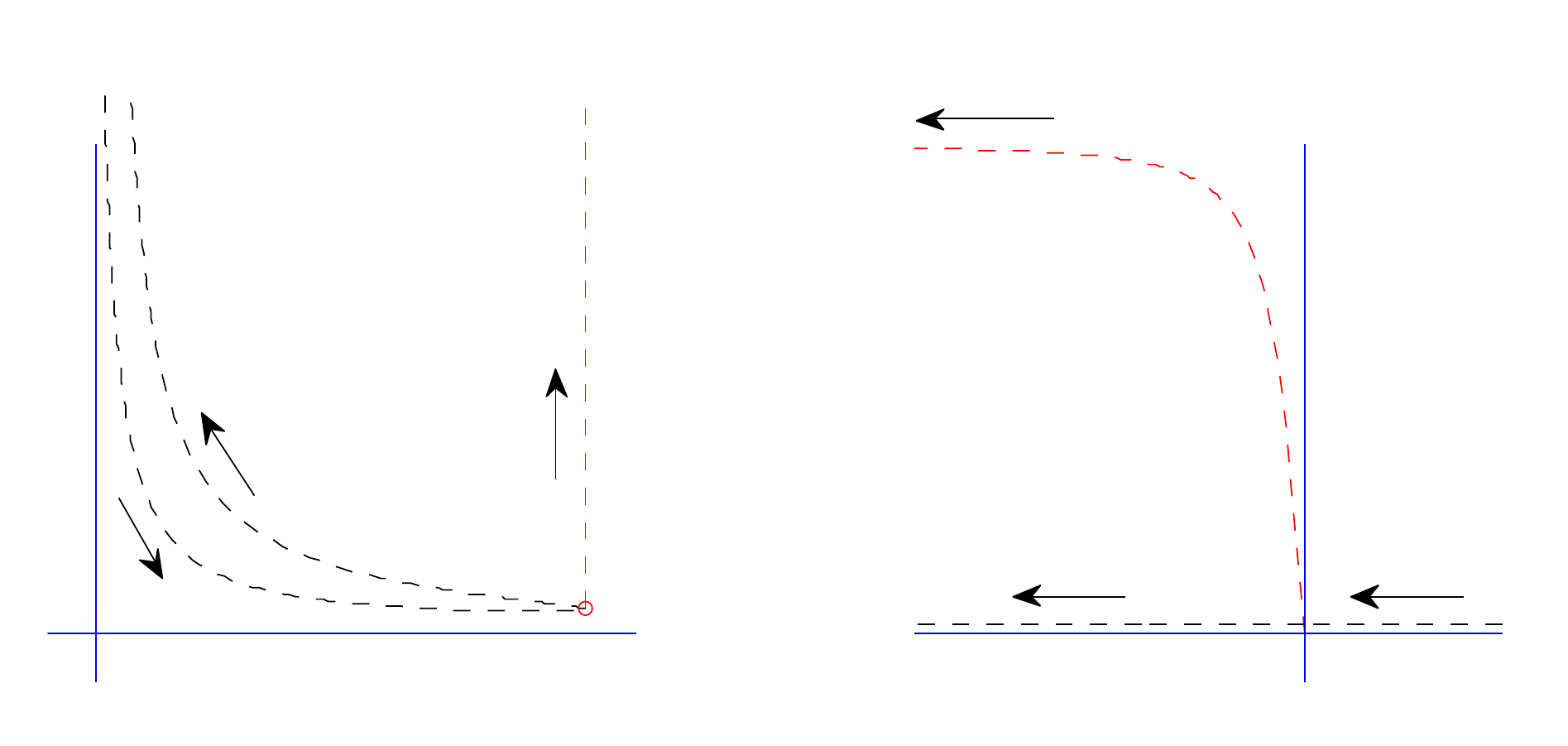}}} 
\put(50,80){$\alpha$}
\put(53,82){\circle{15}}
\put(100,17){\scriptsize $\alpha_{\rm c}$}
\put(26,16){\scriptsize ${\cal B}_1$}
\put(46,39){\scriptsize ${\cal B}_2$}
\put(83,40){\scriptsize ${\cal B}$}
\put(220,80){$\kappa$}
\put(223,82){\circle{15}}
\put(213,21){\scriptsize ${\cal B}_1$}
\put(165,21){\scriptsize ${\cal B}_2$}
\put(153,90){\scriptsize ${\cal B}$}
\end{picture}
\caption{Integration contours for an open coaxial waveguide with a lossy exterior region.
Here ${\cal B}_1$ and ${\cal B}_2$ are the original contours associated with the branch-cut at 
$\alpha_{\rm c}=k_0\sqrt{\mu_{N+1}\epsilon_{N+1}}$,
and ${\cal B}$ the deformed contour. The corresponding contours in the $\kappa$-plane are denoted in the same way
and where $\kappa=\sqrt{\alpha_{\rm c}^2-\alpha^2}$. Note that as ${\cal B}_1$ and ${\cal B}_2$ approach the real line in the
$\kappa$-plane, the contours ${\cal B}_1$ and ${\cal B}_2$ approach each other in the $\alpha$-plane.}
\label{fig:matfig70}
\end{figure}
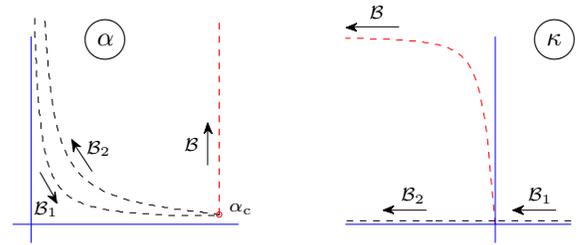

It will be shown in section \ref{sect:Largeargument} that Jordan's lemma \cite{Arfken+Weber2001} 
applies for the integral defined in \eqref{eq:intFdef} and \eqref{eq:Fdef}. 
For $z>0$, the integration contour can therefore be closed in the upper half-plane,
and the current can be written as
\begin{equation}\label{eq:intfsum}
I(\omega,z)=\sum_{p=1}^{\infty}I_p(\omega,z)+I_{\rm br}(\omega,z),
\end{equation}
where $I_p(\omega,z)$ denotes the $\textrm{TM}_{0p}$ modal contributions from the residues at $\alpha_p$ 
\begin{equation}\label{eq:intpdef}
I_p(\omega,z)=\oint_{{\cal C}_p}F(\alpha)\eu^{\iu\alpha z}\diff\alpha,
\end{equation}
where ${\cal C}_p$ is a sufficiently small closed contour containing only the single pole $\alpha_p$, and where
\begin{equation}\label{eq:intbrdef1}
I_{\rm br}(\omega,z)=\int_{{\cal B}_1+{\cal B}_2}F(\alpha)\eu^{\iu\alpha z}\diff\alpha
\end{equation}
is the contribution from the branch-cut. Here 
${\cal B}_1$ and ${\cal B}_2$ are the original contours associated with the branch-cut as depicted in Fig.~\ref{fig:matfig70}. 

The contours in Fig.~\ref{fig:matfig70} have been plotted for a slightly lossy exterior region with $\Im\alpha_{\rm c}>0$,
and in the lossless case when $\Im\alpha_{\rm c}\rightarrow 0$ the contours ${\cal B}_1$ and ${\cal B}_2$ will approach an L-shaped form
in the $\alpha$-plane.
The contribution from the branch-cut can also be expressed as
\begin{equation}
I_{\rm br}(\omega,z)=\int_{{\cal B}_2}q(\alpha)\eu^{\iu\alpha z}\diff\alpha,
\end{equation}
where the function $q(\alpha)$ is defined by the discontinuity at the branch-cut
\begin{equation}\label{eq:qdef1}
q(\alpha)=\left. F(\alpha)\right|_{{\cal B}_2}-\left. F(\alpha)\right|_{{\cal B}_1}.
\end{equation}
Since $q(\alpha)$ is analytic except at the branch-point $\alpha_{\rm c}$ and at the poles $\{\alpha_p\}_{p=1}^{\infty}$,
the contour ${\cal B}_2$ can also be deformed to the more simple contour ${\cal B}$ 
\begin{equation}\label{eq:intbrdef2}
I_{\rm br}(\omega,z)=\int_{{\cal B}}q(\alpha)\eu^{\iu\alpha z}\diff\alpha,
\end{equation}
provided that there are no poles between the two contours ${\cal B}_2$ and ${\cal B}$, see Fig.~\ref{fig:matfig70}.
This assumption is indeed realistic for a simple exterior (vacuum) when $\alpha_{\rm c}=k_0$, and all 
modes can be assumed to have a phase speed  ${\rm c}_0k_0/\Re\alpha_p$ which is less then the speed of light in vacuum ${\rm c}_0$.
It is also a realistic approximation if the extinction coefficients $\Im\alpha_p$ of the poles with $\Re\alpha_p<\Re\alpha_{\rm c}$
are much greater than $\Im\alpha_{\rm c}$. Note that the contour ${\cal B}$ is also a path of steepest descent \cite{Olver1997}. 

It is finally commented that the calibration in \eqref{eq:Mdef} can be carried out based on a spectral decomposition
as in \eqref{eq:intfsum}, since the calibration can be made at $z>0$ to satisfy Jordan's lemma, and then 
letting $z\rightarrow 0+$.

\subsection{Characteristic impedance}\label{sect:charZ}

Assume that the propagation constant $\alpha_p$ of the $\textrm{TM}_{0p}$ mode is given
(\ie has been obtained numerically) for a particular wavenumber $k_0$.
To determine the corresponding characteristic impedance $Z_p$, 
the following quasi-static voltage and current waves are used 
\begin{eqnarray}
V_p^+(z) & = & \int_{\rho_1}^{\rho_L}E_{\rho}\diff\rho, \\
I_p^+(z) & = & \int_0^{2\pi}\int_0^{\rho_1}\sigma_1 E_z \rho\diff\rho\diff\phi,
\end{eqnarray}
where $\rho_1$ is the radius of the inner conductor, and $\rho_L$ the radius of the sheath where the voltage is measured.

By using the expression (\ref{eq:Er}) for $E_{\rho}$, the voltage wave can be computed as
\begin{multline}\label{eq:Vpdef}
V_p^+(z) =  \displaystyle \eu^{\iu\alpha_pz}\frac{\iu\alpha_p}{k_0}\sum_{i=2}^{L}\int_{\rho_{i-1}}^{\rho_i}
\left(a_i {\rm J}_0^{\prime}(\kappa_i\rho) + b_i {\rm H}_0^{(1)\prime}(\kappa_i\rho) \right)\diff\rho\\
=\eu^{\iu\alpha_pz}\frac{\iu\alpha_p}{k_0}\sum_{i=2}^{L}\left\{
 \frac{a_i}{\kappa_i}\left({\rm J}_0(\kappa_i\rho_i)-{\rm J}_0(\kappa_i\rho_{i-1})\right) \right. \\
 \left. +  \frac{b_i}{\kappa_i}\left({\rm H}_0^{(1)}(\kappa_i\rho_i)-{\rm H}_0^{(1)}(\kappa_i\rho_{i-1})\right) \right\}.
\end{multline}
The current wave is computed as in \eqref{eq:Iomegaalphadef}, and is given here by
\begin{equation}\label{eq:Ipdef}
I_p^+(z)=\eu^{\iu\alpha_pz}2\pi\sigma_1\frac{1}{k_0}a_1\rho_1{\rm J}_1(\kappa_1\rho_1).
\end{equation}

The coefficients $a_1$, $a_i$ and $b_i$ for $i=2,\ldots,L$ defining the corresponding voltage and 
current waves in (\ref{eq:Vpdef}) and (\ref{eq:Ipdef}) above
can be obtained as a non-trivial solution from the nullspace of the matrix ${\bf A}(\omega,\alpha_p)$ defined in 
(\ref{eq:Matconds1}) through (\ref{eq:Matconds3})  above.
However, it is a very sensitive numerical operation to obtain an eigenvector of the matrix ${\bf A}(\omega,\alpha_p)$ corresponding
to its zero eigenvalue, in particular when the wavenumber $k_0$ is getting large. 
A preconditioning of the numerical problem is therefore highly recommended, 
and one possibility is to proceed as follows:
Let $a_1=1$, and move the corresponding terms to the right-hand side of \eqref{eq:Matconds1}, yielding
\begin{equation}\label{eq:Matconds1b}
\left\{\begin{array}{l}
 \displaystyle a_2\kappa_{2}{\rm J}_0(\kappa_{2}\rho_1)+b_2\kappa_{2}{\rm H}_0^{(1)}(\kappa_{2}\rho_1)
=\kappa_{1}{\rm J}_0(\kappa_1\rho_1), \vspace{0.2cm}\\
\displaystyle a_2\epsilon_{2} {\rm J}_1(\kappa_{2}\rho_1)+b_2\epsilon_{2}{\rm H}_1^{(1)}(\kappa_{2}\rho_1)
=\epsilon_{1} {\rm J}_1(\kappa_1\rho_1),
\end{array}\right.
\end{equation}
and where there is no excitation, \ie  $M(\omega)=0$.
The equation (\ref{eq:Matconds1b}) together with (\ref{eq:Matconds2}) and (\ref{eq:Matconds3}) represent
an $2N\times(2N-1)$ overdetermined system of linear equations
\begin{equation}\label{eq:Bxb}
{\bf B}{\bf x}={\bf b},
\end{equation}
where ${\bf x}$ is a vector containing the unknowns $a_i$ and $b_i$ for $i=2,\ldots,N$ and $b_{N+1}$.
The system (\ref{eq:Bxb}) can be solved in a least squares sense using a pseudo-inverse \cite{Kelley1995,Greenbaum1997}.
However, this is not sufficient when the problem is severely ill-conditioned. A Jacobi preconditioning \cite{Kelley1995,Greenbaum1997}
is incorporated by instead solving the problem
\begin{equation}\label{eq:BDyb}
{\bf B}{\bf D}^{-1/2}{\bf y}={\bf b}
\end{equation}
where ${\bf y}={\bf D}^{1/2}{\bf x}$ and where ${\bf D}$ is a diagonal matrix with the same diagonal elements as
the matrix ${\bf B}^{\rm H}{\bf B}$. The inversion of (\ref{eq:BDyb}) is well-conditioned since the matrix
$({\bf B}{\bf D}^{-1/2})^{\rm H}{\bf B}{\bf D}^{-1/2}$ has unit diagonal.
The characteristic impedance is then finally obtained as
\begin{equation}
Z_p=\frac{V_p^+(z)}{I_p^+(z)}.
\end{equation}

\subsection{Exponential scaling for stable computations}\label{sect:expscale}
With the power cable application at hand,
simple Perfectly Electrically Conducting (PEC) boundary conditions do not give accurate modeling at the relatively low frequencies
of interest since the skin-depth is large and fields do penetrate the metallic layers. 
Hence, it is crucial to model the fields inside the metals.
However, this requirement gives rise to numerical issues. 
The arguments of the Bessel functions are given by quantities of the form $\kappa_i\rho$ where the transverse wavenumber
is given by
\begin{equation}
\kappa_i=\sqrt{k_0^2\mu_i\epsilon_{{\rm r}i}+\iu k_0\mu_i\sigma_i\eta_0-\alpha^2}
\end{equation}
for a conductive material. Hence, if either of the parameters $\sigma_i$, $\mu_i$ and/or $\epsilon_{{\rm r}i}$ are very large
the corresponding transverse wavenumber can become excessively large and cause numerical problems due to the exponential
growth of Bessel functions, even at low frequencies.
This issue can be remedied by incorporating an exponential scaling of Bessel functions.

The following scaled versions are needed here
\begin{equation}
\left\{\begin{array}{l}
{\rm H}_m^{(1)}(\zeta)=\eu^{\iu \zeta}\tilde{\rm H}_m^{(1)}(\zeta), \vspace{0.2cm} \\
{\rm J}_m(\zeta)=\eu^{-\iu \zeta}\tilde{\rm J}_m(\zeta),
\end{array}\right.
\end{equation}
where $\tilde{\rm H}_m^{(1)}(\zeta)$ is available in numerical software such as MATLAB, and 
$\tilde{\rm J}_m(\zeta)$ can be generated as $\tilde{\rm J}_m(\zeta)=(\eu^{\iu 2\zeta}\tilde{\rm H}_m^{(1)}(\zeta)+\tilde{\rm H}_m^{(2)}(\zeta))/2$ 
and where $\tilde{\rm H}_m^{(2)}(\zeta)$ is the similarly scaled Hankel function of the second kind, 
\ie ${\rm H}_m^{(2)}(\zeta)=\eu^{-\iu \zeta}\tilde{\rm H}_m^{(2)}(\zeta)$.

By introducing the following coefficient substitutions
\begin{equation}\label{eq:coeffsubst1}
\left\{\begin{array}{l}
\tilde{a}_i=a_i\eu^{-\iu \kappa_i\rho_i}, \quad i=1,\ldots,N, \vspace{0.2cm} \\
\tilde{b}_i=b_i\eu^{\iu \kappa_i\rho_i},  \quad i=2,\ldots,N, \vspace{0.2cm} \\
\tilde{b}_{N+1}=b_{N+1}\eu^{\iu \kappa_{N+1}\rho_N},
\end{array}\right.
\end{equation}
the following substitutions can be made in \eqref{eq:Matconds1} through \eqref{eq:Matconds3},
as well as in \eqref{eq:Vpdef} and \eqref{eq:Ipdef}
\begin{equation}\label{eq:coeffsubst2}
\left\{\begin{array}{l}
a_i{\rm J}_m(\kappa_i\rho_i)=\tilde a_i\tilde{\rm J}_m(\kappa_i\rho_i), \quad i=1,\ldots,N, \vspace{0.2cm} \\
b_i{\rm H}_m^{(1)}(\kappa_i\rho_i)=\tilde b_i\tilde{\rm H}_m^{(1)}(\kappa_i\rho_i), \quad i=2,\ldots,N, \vspace{0.2cm} \\
b_{N+1}{\rm H}_m^{(1)}(\kappa_{N+1}\rho_N)=\tilde b_{N+1}\tilde{\rm H}_m^{(1)}(\kappa_{N+1}\rho_N), 
\end{array}\right.
\end{equation}
and
\begin{equation}\label{eq:coeffsubst3}
\left\{\begin{array}{l}
a_{i+1}{\rm J}_m(\kappa_{i+1}\rho_i)=\tilde a_{i+1}\eu^{\iu\kappa_{i+1}d_{i+1}}\tilde{\rm J}_m(\kappa_{i+1}\rho_i), \vspace{0.2cm} \\
b_{i+1}{\rm H}_m^{(1)}(\kappa_{i+1}\rho_i)=\tilde b_{i+1}\eu^{-\iu\kappa_{i+1}d_{i+1}}\tilde{\rm H}_m^{(1)}(\kappa_{i+1}\rho_i), 
\end{array}\right.
\end{equation}
where $d_{i+1}=\rho_{i+1}-\rho_i$ and $i=1,\ldots,N-1$. The factors of exponential growth have now been 
reduced to the factors $\eu^{-\iu\kappa_{i+1}d_{i+1}}$ where the increments $d_{i+1}$ are small. This will greatly
increase the range of frequencies that can be used in a numerical implementation based on the present formulation.
It is finally noted that fundamental combinations such as $a_1{\rm J}_1(\kappa_1\rho_1)$ used
in \eqref{eq:Iomegaalphadef} is left invariant to the scaling.

\section{Recursive computations}\label{sect:recursive}
The determinants $\det{\bf A}(\omega,\alpha)$ and $\det{\bf B}(\omega,\alpha)$ used to define the function $F(\alpha)$
in \eqref{eq:Fdef} can be computed recursively as outlined in the appendix \ref{sect:recdet}.
Hence, based on \eqref{eq:finaldet} together with $x_N^4=\kappa_{N+1}{\rm H}_0^{(1)}(\kappa_{N+1}\rho_N)$ and 
$y_N^4=\epsilon_{N+1}{\rm H}_1^{(1)}(\kappa_{N+1}\rho_N)$
the function $F(\alpha)$ in \eqref{eq:Fdef} is given by
\begin{multline}\label{eq:Fdef2}
F(\alpha)=M(\omega)\sigma_1\rho_1{\rm J}_1(\kappa_1\rho_1)\\
\times\frac{\epsilon_{N+1}{\rm H}_1^{(1)}(\kappa_{N+1}\rho_N)\bar{f}_N-\kappa_{N+1}{\rm H}_0^{(1)}(\kappa_{N+1}\rho_N)\bar{g}_N}
{\epsilon_{N+1}{\rm H}_1^{(1)}(\kappa_{N+1}\rho_N){f}_N-\kappa_{N+1}{\rm H}_0^{(1)}(\kappa_{N+1}\rho_N){g}_N},
\end{multline}
where the auxiliary determinants $f_N$, $g_N$ $\bar{f}_N$ and $\bar{g}_N$ are given recursively by
\begin{equation}\label{eq:figidef}
\left\{\begin{array}{l}
f_1=-\kappa_1{\rm J}_0(\kappa_1\rho_1), \vspace{0.2cm} \\
g_1=-\epsilon_1{\rm J}_1(\kappa_1\rho_1),
\end{array}\right.
\quad \left\{\begin{array}{l}
f_i=A_if_{i-1}+B_ig_{i-1}, \vspace{0.2cm} \\
g_i=C_if_{i-1}+D_ig_{i-1},
\end{array}\right.
\end{equation}
and
\begin{equation}\label{eq:barfigidef}
\left\{\begin{array}{l}
\bar{f}_1=1, \vspace{0.2cm} \\
\bar{g}_1=0,
\end{array}\right.
\quad \left\{\begin{array}{l}
\bar{f}_i=A_i\bar{f}_{i-1}+B_i\bar{g}_{i-1}, \vspace{0.2cm} \\
\bar{g}_i=C_i\bar{f}_{i-1}+D_i\bar{g}_{i-1},
\end{array}\right.
\end{equation}
and where $i=2,\ldots,N$, as in \eqref{eq:recdef}.
The parameters $A_i$, $B_i$, $C_i$ and $D_i$ are defined in \eqref{eq:ABCDdef}, and are
given here explicitly as
\begin{equation}\label{eq:ABCDdef2}
\left\{\begin{array}{l}
A_i(\kappa_i)=\kappa_i\epsilon_i\left[{\rm J}_0(\kappa_i\rho_i){\rm H}_1^{(1)}(\kappa_i\rho_{i-1}) \right.  \\
\hspace{4cm}\left. -{\rm H}_0^{(1)}(\kappa_i\rho_i){\rm J}_1(\kappa_i\rho_{i-1}) \right],  \\
B_i(\kappa_i)=\kappa_i^2\left[{\rm H}_0^{(1)}(\kappa_i\rho_i){\rm J}_0(\kappa_i\rho_{i-1}) \right.  \\
\hspace{4cm}\left. -{\rm J}_0(\kappa_i\rho_i){\rm H}_0^{(1)}(\kappa_i\rho_{i-1}) \right],  \\
C_i(\kappa_i)=\epsilon_i^2\left[{\rm J}_1(\kappa_i\rho_i){\rm H}_1^{(1)}(\kappa_i\rho_{i-1}) \right.  \\
\hspace{4cm}\left. -{\rm H}_1^{(1)}(\kappa_i\rho_i){\rm J}_1(\kappa_i\rho_{i-1}) \right], \vspace{0.2cm} \\
D_i(\kappa_i)=\kappa_i\epsilon_i\left[{\rm H}_1^{(1)}(\kappa_i\rho_i){\rm J}_0(\kappa_i\rho_{i-1}) \right.  \\
\hspace{4cm}\left. -{\rm J}_1(\kappa_i\rho_i){\rm H}_0^{(1)}(\kappa_i\rho_{i-1}) \right],
\end{array}\right.
\end{equation}
where $i=2,\ldots,N$, and where it has been emphasized that they each depend only on the single complex variable $\kappa_i$.

It is noted that in the computation of the dispersion function $\det{\bf A}(\omega,\alpha)$ corresponding to a
multilayered open waveguide, the auxiliary determinants $f_i$ and $g_i$ can be interpreted as intermediate dispersion functions
on their own right, corresponding to a termination with PEC (Perfect Electric Conducting) and PMC (Perfect Magnetic Conducting) 
boundary conditions at radius $\rho_i$, respectively.

The expression \eqref{eq:Fdef2} and the recursive formulation of the auxiliary determinants \eqref{eq:figidef}
and \eqref{eq:barfigidef} are well suited for asymptotic analysis which is the topic of the next section. Hence, for analytic convenience,
the expressions \eqref{eq:Fdef2} through \eqref{eq:ABCDdef2} are given here without the exponential scalings.

From a computational point of view the recursive computation of the determinants in \eqref{eq:figidef}
and \eqref{eq:barfigidef} is numerically efficient since the recursion can operate directly (in parallel)
on an array of data in the complex $\alpha$-plane. The exponential scalings defined in \eqref{eq:coeffsubst1}
through \eqref{eq:coeffsubst3} can readily be implemented by making the corresponding modifications in 
the definition of the recursion based on \eqref{eq:x1234def} and \eqref{eq:y1234def} given in the appendix \ref{sect:recdet}.
In particular, let $\tilde{f}_N$, $\tilde{g}_N$, $\tilde{\bar{f}}_N$ and $\tilde{\bar{g}}_N$ denote
the correspondingly scaled determinants. It can then be shown that
\begin{equation}\label{eq:scaledfg}
\left\{\begin{array}{l}
\tilde{f}_N=\eu^{\iu\kappa_1\rho_1}f_N, \vspace{0.2cm} \\
\tilde{g}_N=\eu^{\iu\kappa_1\rho_1}g_N,
\end{array}\right.
\quad
\left\{\begin{array}{l}
\tilde{\bar{f}}_N=\bar{f}_N, \vspace{0.2cm} \\
\tilde{\bar{g}}_N=\bar{g}_N,
\end{array}\right.
\end{equation}
which leaves \eqref{eq:Fdef2} invariant to the scaling, as should be expected.

\section{Asymptotic analysis}
\subsection{Small argument asymptotics and analytic properties}\label{sect:analytprop}
The following definitions and small argument asymptotics will be useful throughout the analysis,
see \cite{Olver+etal2010}. The regular Bessel functions of order $m=0,1$ are even and odd analytic functions,
respectively, and with the following asymptotics
\begin{equation}\label{eq:J0J1def}
\left\{\begin{array}{l}
{\rm J}_0(\zeta)=1-\frac{1}{4}\zeta^2+{\cal O}\{\zeta^4\}, \vspace{0.2cm} \\
{\rm J}_1(\zeta)  = \frac{1}{2}\zeta+{\cal O}\{\zeta^3\},
\end{array}\right.
\end{equation}
where ${\cal O}\{\cdot\}$  denotes the big ordo \cite{Olver1997,Olver+etal2010}.
The Neumann function of order $m=0$ is given by
\begin{equation}\label{eq:Y0def}
{\rm Y}_0(\zeta)=\frac{2}{\pi}\ln\frac{\zeta}{2}{\rm J}_0(\zeta)+B(\zeta),
\end{equation}
where $B(\zeta)$ is an even analytic function with asymptotics
\begin{equation}\label{eq:Bzdef}
B(\zeta)=\frac{2}{\pi}\gamma+\frac{1-\gamma}{2\pi}\zeta^2+{\cal O}\{\zeta^4\},
\end{equation}
and where $\gamma$ is Euler's constant.
The Neumann function of order $m=1$ is similarly given by
\begin{equation}\label{eq:Y1def}
{\rm Y}_1(\zeta)=\frac{2}{\pi}\ln\frac{\zeta}{2}{\rm J}_1(\zeta)+C(\zeta),
\end{equation}
where $C(\zeta)$ is an odd meromorphic function with asymptotics
\begin{equation}\label{eq:Czdef}
C(\zeta)=-\frac{2}{\pi}\frac{1}{\zeta}+\frac{2\gamma-1}{2\pi}\zeta+{\cal O}\{\zeta^3\}.
\end{equation}
The Hankel functions of the first and second kind are defined by ${\rm H}_m^{(1)}(\zeta)={\rm J}_m(\zeta)+\iu{\rm Y}_m(\zeta)$
and ${\rm H}_m^{(2)}(\zeta)={\rm J}_m(\zeta)-\iu{\rm Y}_m(\zeta)$, respectively.
From the definitions above it follows that
\begin{equation}\label{eq:Hankelexpr}
\left\{\begin{array}{l}
{\rm H}_0^{(1)}(\zeta)={\rm J}_0(\zeta)+\iu B(\zeta)+\iu\frac{2}{\pi}\ln\frac{\zeta}{2}{\rm J}_0(\zeta), \vspace{0.2cm} \\
{\rm H}_0^{(2)}(\zeta)={\rm J}_0(\zeta)-\iu B(\zeta)-\iu\frac{2}{\pi}\ln\frac{\zeta}{2}{\rm J}_0(\zeta), \vspace{0.2cm} \\
{\rm H}_1^{(1)}(\zeta)={\rm J}_1(\zeta)+\iu C(\zeta)+\iu\frac{2}{\pi}\ln\frac{\zeta}{2}{\rm J}_1(\zeta),\vspace{0.2cm}  \\
{\rm H}_1^{(2)}(\zeta)={\rm J}_1(\zeta)-\iu C(\zeta)-\iu\frac{2}{\pi}\ln\frac{\zeta}{2}{\rm J}_1(\zeta).
\end{array}\right.
\end{equation}

Based on the definitions in \eqref{eq:Hankelexpr}, the parameters $A_i$, $B_i$, $C_i$ and $D_i$
defined in \eqref{eq:ABCDdef2} can be expressed as
\begin{equation}\label{eq:ABCDdef3}
\left\{\begin{array}{l}
A_i(\kappa_i)=\iu\kappa_i\epsilon_i\left[{\rm J}_0(\kappa_i\rho_i)C(\kappa_i\rho_{i-1}) \right.  \\
\left. -{\rm J}_1(\kappa_i\rho_{i-1})B(\kappa_i\rho_{i})
+{\rm J}_0(\kappa_i\rho_i){\rm J}_1(\kappa_i\rho_{i-1})\frac{2}{\pi}\ln\frac{\rho_{i-1}}{\rho_i} \right],  \vspace{0.2cm}\\
B_i(\kappa_i)=\iu\kappa_i^2\left[{\rm J}_0(\kappa_i\rho_{i-1})B(\kappa_i\rho_i) \right.  \\
\left. -{\rm J}_0(\kappa_i\rho_{i})B(\kappa_i\rho_{i-1})
+{\rm J}_0(\kappa_i\rho_{i-1}){\rm J}_0(\kappa_i\rho_{i})\frac{2}{\pi}\ln\frac{\rho_{i}}{\rho_{i-1}}  \right],  \vspace{0.2cm}\\
C_i(\kappa_i)=\iu\epsilon_i^2\left[{\rm J}_1(\kappa_i\rho_i)C(\kappa_i\rho_{i-1}) \right.  \\
\left. -{\rm J}_1(\kappa_i\rho_{i-1})C(\kappa_i\rho_{i})
+{\rm J}_1(\kappa_i\rho_i){\rm J}_1(\kappa_i\rho_{i-1})\frac{2}{\pi}\ln\frac{\rho_{i-1}}{\rho_i} \right], \vspace{0.2cm} \\
D_i(\kappa_i)=\iu\kappa_i\epsilon_i\left[{\rm J}_0(\kappa_i\rho_{i-1})C(\kappa_i\rho_i) \right.  \\
\left. -{\rm J}_1(\kappa_i\rho_{i})B(\kappa_i\rho_{i-1})
+{\rm J}_0(\kappa_i\rho_{i-1}){\rm J}_1(\kappa_i\rho_{i})\frac{2}{\pi}\ln\frac{\rho_{i}}{\rho_{i-1}} \right],\vspace{0.2cm} \\
\end{array}\right.
\end{equation}
where $i=2,\ldots,N$. It is observed that the logarithmic singularities of the Hankel functions 
vanish, and the parameters $A_i$, $B_i$, $C_i$ and $D_i$ are even analytic functions
in the complex variable $\kappa_i$, with asymptotics
\begin{equation}\label{eq:ABCDdef4}
\left\{\begin{array}{l}
A_i(\kappa_i)={\cal O}\{1\},  \vspace{0.2cm}\\
B_i(\kappa_i)={\cal O}\{\kappa_i^2\},  \vspace{0.2cm}\\
C_i(\kappa_i)= {\cal O}\{1\},\vspace{0.2cm} \\
D_i(\kappa_i)={\cal O}\{1\}.
\end{array}\right.
\end{equation}

From the recursions \eqref{eq:figidef} and \eqref{eq:barfigidef} follows that the determinants
$f_N$, $g_N$, $\bar{f}_N$ and $\bar{g}_N$ are even analytic functions in each variable
$\kappa_i$ for $i=2,\ldots,N$. Furthermore, the two determinants $f_N$ and $g_N$
are odd analytic functions in the variable $\kappa_1$. In combination with the odd Bessel function
${\rm J}_1(\kappa_1\rho_1)$ in \eqref{eq:Fdef2}, it follows that $F(\alpha)$ is an even analytic
function in a neighborhood of $\kappa_i=0$, for each variable $\kappa_i$ for $i=1,\ldots,N$. Hence, it is concluded that $F(\alpha)$ is a meromorphic
function in the complex variable $\alpha$ with poles at $\{\alpha_p\}_{p=1}^{\infty}$ and 
one single branch-point at $\alpha_{\rm c}=k_0\sqrt{\mu_{N+1}\epsilon_{N+1}}$,
corresponding to the wavenumber of the exterior domain, and where $\kappa_{N+1}=\sqrt{\alpha_{\rm c}^2-\alpha^2}$.

\subsection{Large argument asymptotics and Jordan's lemma}\label{sect:Largeargument}
For large values of $\alpha$ the transverse wavenumbers behave asymptotically as
$\kappa_i=\sqrt{k_0^2\mu_i\epsilon_i-\alpha^2}\sim\pm\iu\alpha$ where the branch of the square root has been chosen so that
$\Im\kappa_i\geq 0$.
Based on the following large argument asymptotics of the Hankel functions of the first and second kind, and of the 
regular Bessel functions 
\begin{equation}\label{eq:Besselas}
\left\{\begin{array}{l}
{\rm H}_m^{(1)}(\zeta)\sim\sqrt{\frac{2}{\pi \zeta}}\eu^{\iu(\zeta-\frac{1}{2}m\pi-\frac{1}{4}\pi)}, \vspace{0.2cm} \\
{\rm H}_m^{(2)}(\zeta)\sim\sqrt{\frac{2}{\pi \zeta}}\eu^{-\iu(\zeta-\frac{1}{2}m\pi-\frac{1}{4}\pi)}, \vspace{0.2cm} \\
{\rm J}_m(\zeta)\sim\sqrt{\frac{2}{\pi \zeta}}\cos(\zeta-\frac{1}{2}m\pi-\frac{1}{4}\pi),
\end{array}\right.
\end{equation}
the following asymptotic properties of the parameters $A_i$, $B_i$, $C_i$ and $D_i$ can be derived based on \eqref{eq:ABCDdef2}
\begin{equation}\label{eq:ABCDas}
\left\{\begin{array}{l}
A_i(\kappa_i)\sim\displaystyle -2\iu\epsilon_i\frac{1}{\pi\sqrt{\rho_i\rho_{i-1}}}\cos\kappa_id_i, \vspace{0.2cm} \\
B_i(\kappa_i)\sim\displaystyle 2\iu\kappa_i\frac{1}{\pi\sqrt{\rho_i\rho_{i-1}}}\sin\kappa_id_i, \vspace{0.2cm} \\
C_i(\kappa_i)\sim\displaystyle -2\iu\frac{\epsilon_i^2}{\kappa_i}\frac{1}{\pi\sqrt{\rho_i\rho_{i-1}}}\sin\kappa_id_i, \vspace{0.2cm} \\
D_i(\kappa_i)\sim\displaystyle -2\iu\epsilon_i\frac{1}{\pi\sqrt{\rho_i\rho_{i-1}}}\cos\kappa_id_i,
\end{array}\right.
\end{equation}
where $d_i=\rho_i-\rho_{i-1}$ and $i=2,\ldots,N$.

A detailed study of the recursion in \eqref{eq:figidef} and \eqref{eq:barfigidef} together with the asymptotic properties in \eqref{eq:ABCDas}
shows that
\begin{equation}\label{eq:fNgNas}
\left\{\begin{array}{l}
\displaystyle f_N\sim f_1 P_N+g_1\alpha Q_N, \vspace{0.2cm} \\
\displaystyle g_N\sim f_1 \frac{1}{\alpha}R_N+g_1 S_N,
\end{array}\right.
\quad
\left\{\begin{array}{l}
\displaystyle \bar{f}_N\sim P_N, \vspace{0.2cm} \\
\displaystyle \bar{g}_N\sim \frac{1}{\alpha}R_N,
\end{array}\right.
\end{equation}
where $P_N$, $Q_N$, $R_N$ and $S_N$ are $N-1$ order polynomials where each term consists of 
product combinations of the type
\begin{equation}
\left(\begin{array}{l}
\cos\kappa_2d_2 \vspace{0.2cm} \\
\sin\kappa_2d_2
\end{array}\right) \cdots
\left(\begin{array}{l}
\cos\kappa_Nd_N \vspace{0.2cm} \\
\sin\kappa_Nd_N
\end{array}\right),
\end{equation}
with one cosine or sine factor for each layer index $i=2,\ldots,N$.

A detailed study of \eqref{eq:Fdef2} based on the asymptotics given in \eqref{eq:Besselas} and \eqref{eq:fNgNas}
shows that the function $F(\alpha)$ has the large argument asymptotics
\begin{equation}
F(\alpha)={\cal O}\left\{\frac{1}{\alpha}\right\}.
\end{equation}
It is noted that the factors $P_N$, $Q_N$, $R_N$ and $S_N$ are either bounded and oscillates, such as \eg on the contour ${\cal B}$
as depicted in Fig.~\ref{fig:matfig70}, or they are all growing with the same exponential order.
Since $F(\alpha)\rightarrow 0$ as $\alpha\rightarrow\infty$, Jordan's lemma \cite{Arfken+Weber2001} 
applies for the integral defined in \eqref{eq:intFdef}, and when $z>0$ the integration contour can be closed in the upper half-plane. 

Finally, it is interesting to note the following: If instead an electrical frill generator is used with surface current
density $\vec{J}_{\rm S}=J(\omega)\delta(z)\hat{\vec{z}}$, the situation can be analyzed similarly as above  
using $\bar{f}_1=0$ and $\bar{g}_1=1$. The result is that this time the current in the inner conductor has
the large argument asymptotics $F(\alpha)={\cal O}\{1\}$.
A higher degree of regularity must therefore be imposed on the spatial 
excitation in order to satisfy the prerequisites for Jordan's lemma. 
Hence, it is preferable to use the magnetic frill generator due to faster spectral decay for large values of $\alpha$.

\subsection{Behaviour at the branch-point and asymptotic approximations}

Based on \eqref{eq:Fdef2}, the function $q(\alpha)$ defined in \eqref{eq:qdef1} can be written
\begin{multline}\label{eq:qdef2}
q(\alpha)=\left. F(\alpha)\right|_{{\cal B}_2}-\left. F(\alpha)\right|_{{\cal B}_1} \\
=\frac{M(\omega)4\iu\sigma_1\frac{\rho_1}{\rho_N}{\rm J}_1(\kappa_1\rho_1)\epsilon_{N+1}(\bar{f}_Ng_N-f_N\bar{g}_N)}
{\pi D_1(\alpha)D_2(\alpha)},
\end{multline}
where
\begin{equation}
D_1(\alpha)=\epsilon_{N+1}{\rm H}_1^{(1)}(\kappa_{N+1}\rho_N)f_N-\kappa_{N+1}{\rm H}_0^{(1)}(\kappa_{N+1}\rho_N)g_N,
\end{equation}
and
\begin{equation}
D_2(\alpha)=\epsilon_{N+1}{\rm H}_1^{(2)}(\kappa_{N+1}\rho_N)f_N-\kappa_{N+1}{\rm H}_0^{(2)}(\kappa_{N+1}\rho_N)g_N.
\end{equation}
To derive \eqref{eq:qdef2}, the following identities have been used to account for the discontinuity over the branch-cut,
where $\kappa_{N+1}|_{{\cal B}_1}=-\kappa_{N+1}|_{{\cal B}_2}$ and
\begin{equation}
{\rm H}_m^{(1)}(\zeta)|_{{\cal B}_1}={\rm H}_m^{(1)}(-\zeta)|_{{\cal B}_2}
=-(-1)^m{\rm H}_m^{(2)}(\zeta)|_{{\cal B}_2},
\end{equation}
as well as the Wronskian 
${\rm H}_0^{(2)}(\zeta){\rm H}_1^{(1)}(\zeta)-{\rm H}_0^{(1)}(\zeta){\rm H}_1^{(2)}(\zeta)=-4\iu/(\pi \zeta)$, where
$\zeta=\kappa_{N+1}\rho_N$.

The function $q(\alpha)$ is analytic in any simply connected open set that intersects the contour ${\cal B}$ and does not contain
the branch-point $\alpha_{\rm c}$ or any of the poles $\{\alpha_p\}_{p=1}^{\infty}$. The function is not analytic in any neighborhood of $\alpha_{\rm c}$.
However, a detailed analysis of $q(\alpha)$ based on \eqref{eq:qdef2} shows that $q(\alpha)$ and $q^{\prime}(\alpha)$ 
are continuous at $\alpha_{\rm c}$, and $q^{\prime\prime}(\alpha)$ has a logarithmic singularity at $\alpha_{\rm c}$.
In particular, the analysis shows that
\begin{equation}
\left\{\begin{array}{l}
q(\alpha_{\rm c})=0, \vspace{0.2cm} \\
\displaystyle q^\prime(\alpha_{\rm c})=-\frac{\iu 2\pi\alpha_{\rm c} M(\omega)\sigma_1\rho_1\rho_N{\rm J}_1(\kappa_1\rho_1)
(\bar{f}_Ng_N-f_N\bar{g}_N)}{\epsilon_{N+1}f_N^2}, \vspace{0.2cm} \\
q^{\prime\prime}(\alpha)=A\ln(-\iu(\alpha-\alpha_{\rm c}))+r(\alpha),
\end{array}\right.
\end{equation}
where the constant $A$ is given by
\begin{equation}
A=-q^\prime(\alpha_{\rm c})\frac{\epsilon_{N+1}f_N-2g_N/\rho_N}{\epsilon_{N+1}f_N}2\alpha_{\rm c}\rho_N^2,
\end{equation}
and $r(\alpha)$ is a continuous rest term. The rest term is assumed to be bounded on the contour ${\cal B}$ as
\begin{equation}\label{eq:restterm}
\left|r(\alpha)\right|\leq M,\quad \alpha\in{\cal B}.
\end{equation}
It is also noted that the large argument asymptotics of $q(\alpha)$ and $q^\prime(\alpha)$ on ${\cal B}$ is 
$q(\alpha)={\cal O}\{\alpha^{-1}\}$ and $q^\prime(\alpha)={\cal O}\{\alpha^{-2}\}$.

A repeated integration by parts can now be carried out to yield
\begin{equation}\label{eq:intbrapprox}
I_{\rm br}(\omega,z)=\int_{{\cal B}}q(\alpha)\eu^{\iu\alpha z}\diff\alpha=I_{\rm br}^{\rm as}(\omega,z)+e(\omega,z),
\end{equation}
where the asymptotic approximation for large $z$ is given by
\begin{equation}\label{eq:intbrasdef}
I_{\rm br}^{\rm as}(\omega,z)=\eu^{\iu\alpha_{\rm c}z}\left[ -\frac{q^\prime(\alpha_{\rm c})}{z^2}-\frac{\iu A}{z^3}(-\gamma-\ln z)   \right],
\end{equation}
and $e(\omega,z)$ is the error term defined by
\begin{equation}\label{eq:edef}
e(\omega,z)=-\frac{1}{z^2}\int_{{\cal B}}r(\alpha)\eu^{\iu\alpha z}\diff\alpha,
\end{equation}
see also \cite{Nordebo+etal2013a,Nilsson+etal2012,Olver1997}.
In order to derive \eqref{eq:intbrapprox} through \eqref{eq:edef}, the following integral has been used
\begin{multline}
\int_{{\cal B}}\ln (-\iu(\alpha-\alpha_{\rm c}))\eu^{\iu\alpha z}\diff\alpha
=\left\{\alpha=\alpha_{\rm c}+\iu t \right\}\\
=\int_{0}^{\infty}\ln t \eu^{\iu(\alpha_{\rm c}+\iu t) z}\iu\diff t=\iu\eu^{\iu\alpha_{\rm c}z}\left(-\frac{\gamma}{z}-\frac{\ln z}{z}\right),
\end{multline}
where the last integral is evaluated by using the unilateral Laplace transform of the function $\ln t$, see \eg \cite{Abramowitz+Stegun1970}.

It can similarly be shown that the error term is bounded by
\begin{equation}\label{eq:eubdef}
\left| e(\omega,z)\right|\leq M\eu^{-\Im \alpha_{\rm c}z}\frac{1}{z^3}=e^{\rm ub}(\omega,z).
\end{equation}
Hence, an upper bound for the branch-cut contribution $I_{\rm br}(\omega,z)$ based on the asymptotic expressions 
\eqref{eq:intbrasdef} and \eqref{eq:eubdef} is given by
\begin{equation}\label{eq:Ibrub}
\left|I_{\rm br}(\omega,z)\right|\leq \left|I_{\rm br}^{\rm as}(\omega,z)\right|+e^{\rm ub}(\omega,z)=I_{\rm br}^{\rm ub}(\omega,z).
\end{equation}

\section{Numerical examples}
\begin{figure}[htb]
\begin{picture}(50,40)
\put(100,0){\makebox(50,40){\includegraphics[width=5cm]{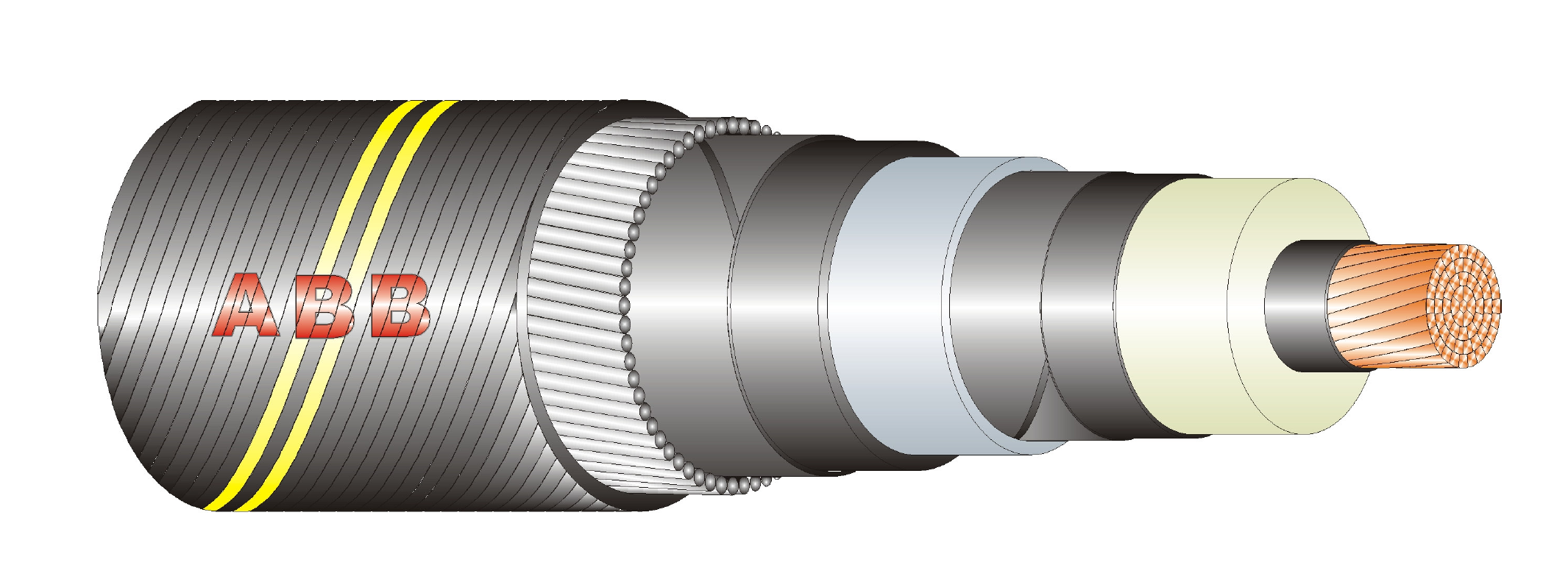}}} 
\end{picture}
\caption{Cross-section of the extruded HVDC sea cable.}
\label{fig:illustration1}
\end{figure}

As a concrete and an industrially relevant numerical example we consider an 82\unit{km} long 200\unit{kV} HVDC 
sea cable as depicted in Fig.~\ref{fig:illustration1}.

The modeling parameters are shown in Table \ref{tab:tab1} where
$\rho_i$ is the radius, $\epsilon_{{\rm r}i}$ the real relative permittivity and $\sigma_i$ the conductivity 
for all layers where $i=1,\ldots,12$. This information is based on data sheets and drawings, but also
on available literature \cite{Gustavsen+etal2005,Heinrich+etal2000,Yang+Zhang2007,Gudmundsdottir+etal2010}. 
It is \eg well known that the real permittivity of the semi-conducting layers (conductor screen and insulation screen) 
can be very large \cite{Gustavsen+etal2005,Heinrich+etal2000,Yang+Zhang2007}, and that the swelling tapes of the conductor binder and
the longitudinal water barrier are usually semi-conducting \cite{Gudmundsdottir+etal2010}. The most crucial parameters at 
low frequencies are the conductivities of the conductor, the lead sheath and the armour which have been modeled 
here for an ambient temperature\footnote{The validation measurements were performed during a very hot summer week in June 2011.}
of 30\degree\unit{C}. It is noted that the conductivity of copper ($\sigma=5.0\cdot 10^7$\unit{S/m} at $20\degree\unit{C}$) has been  
modified with respect to the correct copper area (roughly 91\% for hexagonal close packed threads) and the conductivity of
steel ($\sigma=5.55\cdot 10^6\pi/4$\unit{S/m} at $20\degree\unit{C}$) has been modified with respect to a rectangular close packing of wires.
The lead sheath is assumed to consist of pure lead ($\sigma=4.6\cdot 10^6$\unit{S/m} at $20\degree\unit{C}$) since
the amounts of cadmium and tin are so small that it barely affects the conductivity \cite{Worzyk2009,Cadirli+etal2011}. 
The armour is also slightly magnetic which has been modeled
by using an estimated relative permittivity of $\mu_{10}=40$. All other layers are modeled with $\mu_i=1$.

\begin{table}[htb]\centerline{
\begin{tabular}{||l||l|l|l|}
\cline{1-4} \scriptsize
{\bf Layer} &  \scriptsize $\rho_i$ [\unit{mm}] &  \scriptsize $\epsilon_{{\rm r}i}$ & \scriptsize $\sigma_i$ [\unit{S/m}]      \\
\cline{1-4} \scriptsize 1. Conductor (copper) & \scriptsize $24.3$  & \scriptsize $1$  & \scriptsize $4.81\cdot 10^7$     \\
\cline{1-4} \scriptsize 2. Conductor binder & \scriptsize $24.5$  & \scriptsize $3$  & \scriptsize $1$     \\
\cline{1-4} \scriptsize 3. Conductor screen & \scriptsize $26.1$  & \scriptsize $1000$  & \scriptsize $1$      \\
\cline{1-4} \scriptsize 4. Insulation & \scriptsize $42.0$  & \scriptsize $2.3$  & \scriptsize $0$      \\
\cline{1-4} \scriptsize 5. Insulation screen & \scriptsize $43.2$  & \scriptsize  $1000$  & \scriptsize $1$     \\
\cline{1-4} \scriptsize 6. Longitudinal water barrier & \scriptsize $43.9$  & \scriptsize  $3$  & \scriptsize $1$     \\
\cline{1-4} \scriptsize 7. Lead sheath & \scriptsize $46.9$  & \scriptsize $1$   & \scriptsize $4.43\cdot 10^6$     \\
\cline{1-4} \scriptsize 8. Inner sheath & \scriptsize $49.3$  & \scriptsize $3$   & \scriptsize $0$      \\
\cline{1-4} \scriptsize 9. Bedding & \scriptsize $49.5$  & \scriptsize $3$   & \scriptsize $0$      \\
\cline{1-4} \scriptsize 10. Armour (steel wires) & \scriptsize $54.5$  & \scriptsize $1.4$   & \scriptsize $4.15\cdot 10^6$      \\
\cline{1-4} \scriptsize 11. Outer serving & \scriptsize $58.5$  & \scriptsize $3$   & \scriptsize $0$    \\
\cline{1-4} \scriptsize 12. Exterior region & \scriptsize $\infty$  & \scriptsize $1$   & \scriptsize $0$     \\
\cline{1-4} 
\end{tabular}}
\hspace{5mm}
\caption{Modeling parameters.}\label{tab:tab1}
\end{table}

The exterior region has been modeled as vacuum (or air) as the validation measurements were
performed on a sea cable that was rolled up on shore,
see also \cite{TEAT-7211}. It should also be noted that cable measurements for detecting partial discharges etc.,
often are carried out on land before the cable is put in place at sea.

\begin{figure}[htb]
\begin{picture}(50,120)
\put(100,0){\makebox(50,100){\includegraphics[width=9cm]{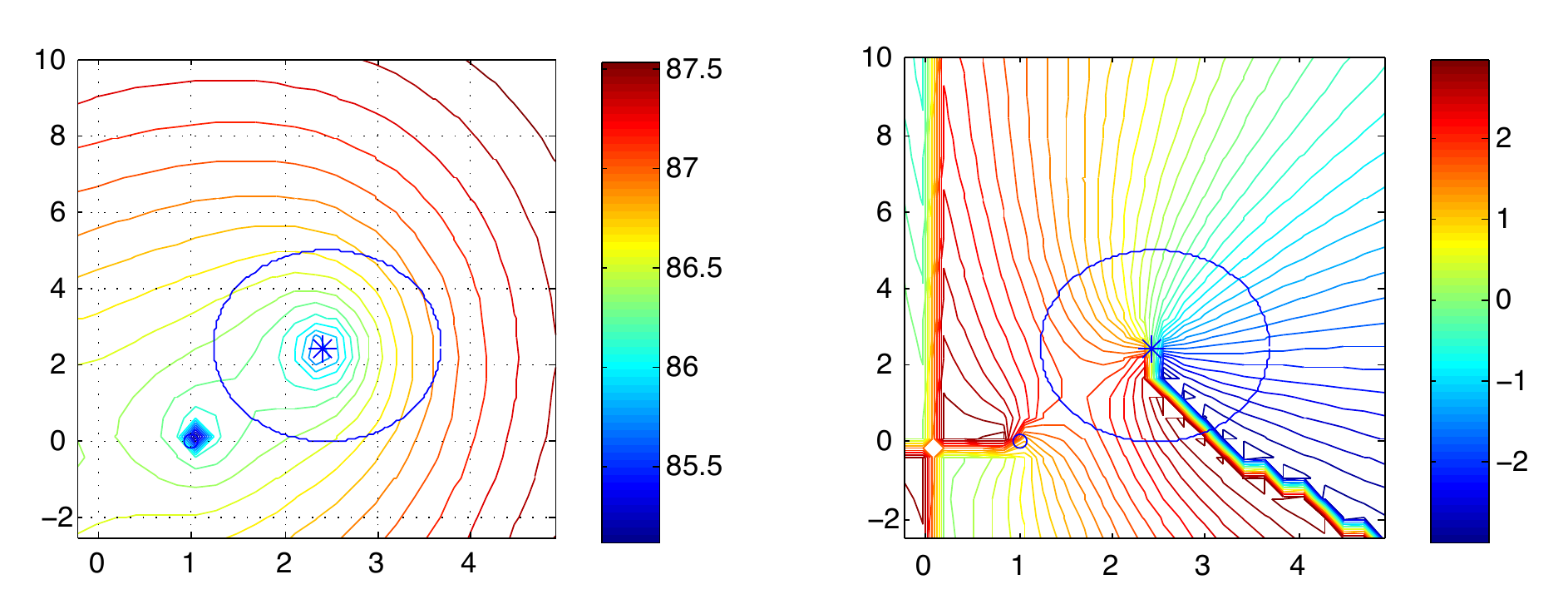}}} 
\put(25,105){\scriptsize a) $\log |\det\bm{A}(\alpha)|$}
\put(0,95){\scriptsize $\Im\alpha\unit{[dB/100km]}$}
\put(42,-2){\scriptsize $\Re\alpha/k_0$}
\put(50,21){\scriptsize ${\cal C}_1$}
\put(153,105){\scriptsize b) $\arg \det\bm{A}(\alpha)\unit{[rad]}$}
\put(136,95){\scriptsize $\Im\alpha\unit{[dB/100km]}$}
\put(178,-2){\scriptsize $\Re\alpha/k_0$}
\put(185,21){\scriptsize ${\cal C}_1$}
\end{picture}
\caption{Illustration of pole search. a) Amplitude of the dispersion function $\det\bm{A}(\alpha)$. 
b) Argument of the dispersion function $\det\bm{A}(\alpha)$.
The complex variable $\alpha$ is scaled as $\Re\alpha/k_0$ (dimensionless) and $\Im\alpha \cdot 2\cdot 10^{6}\log\eu \unit{[dB/100km]}$.
The branch-point $\alpha_{\rm c}=k_0$ is indicated by the circle ``o''. The frequency is $f=150$\unit{Hz} and the contour ${\cal C}_1$ encloses the
pole $\alpha_1$ of the quasi-TEM $\textrm{TM}_{01}$ mode, and which is indicated by the ``*''.}
\label{fig:matfig10}
\end{figure}
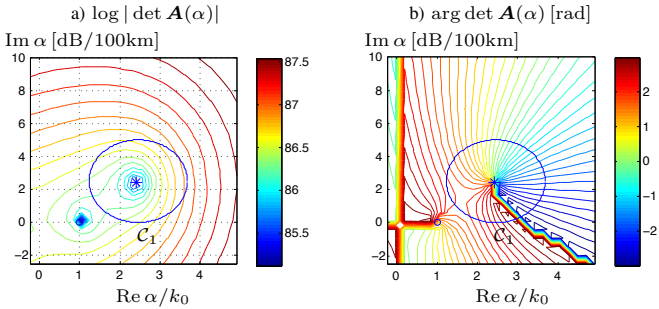
\begin{figure}[htb]
\begin{picture}(50,120)
\put(100,0){\makebox(50,100){\includegraphics[width=9cm]{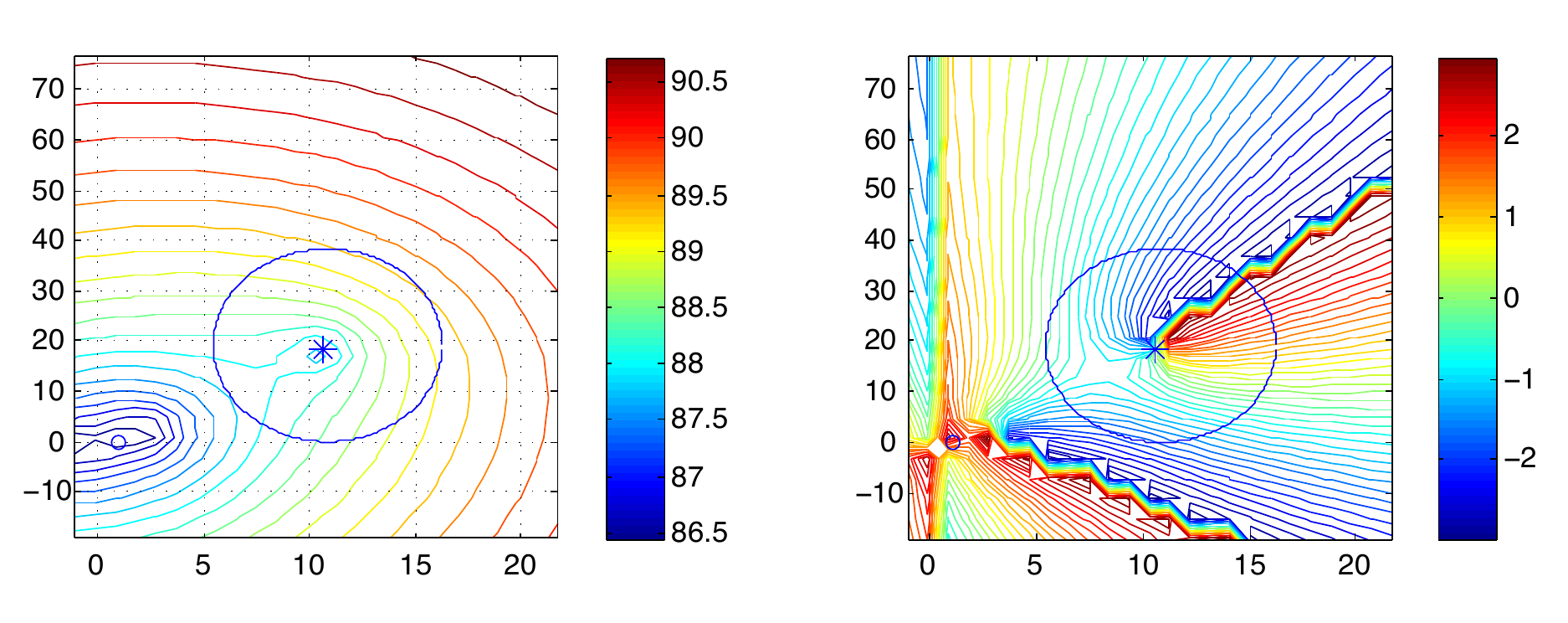}}} 
\put(25,105){\scriptsize a) $\log |\det\bm{A}(\alpha)|$}
\put(0,95){\scriptsize $\Im\alpha\unit{[dB/100km]}$}
\put(42,-2){\scriptsize $\Re\alpha/k_0$}
\put(62,25){\scriptsize ${\cal C}_2$}
\put(153,105){\scriptsize b) $\arg \det\bm{A}(\alpha)\unit{[rad]}$}
\put(136,95){\scriptsize $\Im\alpha\unit{[dB/100km]}$}
\put(178,-2){\scriptsize $\Re\alpha/k_0$}
\put(200,25){\scriptsize ${\cal C}_2$}
\end{picture}
\caption{Illustration of pole search. a) Amplitude of the dispersion function $\det\bm{A}(\alpha)$. 
b) Argument of the dispersion function $\det\bm{A}(\alpha)$.
The complex variable $\alpha$ is scaled as $\Re\alpha/k_0$ (dimensionless) and $\Im\alpha \cdot 2\cdot 10^{6}\log\eu \unit{[dB/100km]}$.
The branch-point $\alpha_{\rm c}=k_0$ is indicated by the circle ``o''. The frequency is $f=150$\unit{Hz} and the contour ${\cal C}_2$ encloses the
pole $\alpha_2$ of the $\textrm{TM}_{02}$ mode, and which is indicated by the ``*''.}
\label{fig:matfig20}
\end{figure}
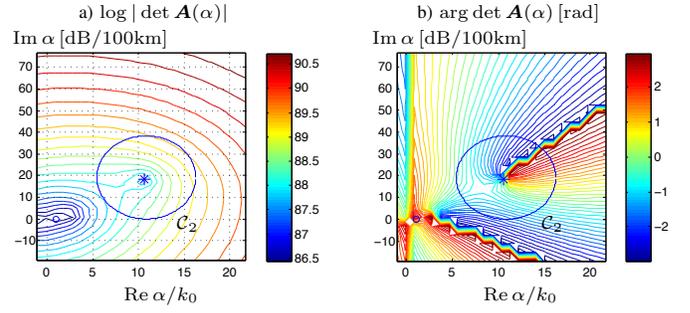

The numerical computations are based on root finding and numerical contour integration.
The poles $\alpha_p$ are found by studying the dispersion equation $\det {\bf A}(\omega,\alpha)=0$,
and computed numerically as
\begin{equation}\label{eq:contour}
\alpha_{p}=\displaystyle\frac{\displaystyle\oint_{{\cal C}_p}\displaystyle\frac{\alpha}{\det{\bf A}(\omega,\alpha)}\diff\alpha}
{\displaystyle\oint_{{\cal C}_p}\displaystyle\frac{1}{\det{\bf A}(\omega,\alpha)}\diff\alpha},
\end{equation}
provided that the closed loop ${\cal C}_p$ is circumscribing the true value $\alpha_{p}$, and that there are no other zeros 
or branch-points of $\det{\bf A}(\omega,\alpha)$ inside the loop.
The pole search is illustrated in Figs.~\ref{fig:matfig10} and \ref{fig:matfig20}
regarding the quasi-TEM $\textrm{TM}_{01}$ mode and the $\textrm{TM}_{02}$ mode at $f=150$\unit{Hz}, respectively.
More details on the zero-finding algorithm can also be found in \cite{TEAT-7211}.

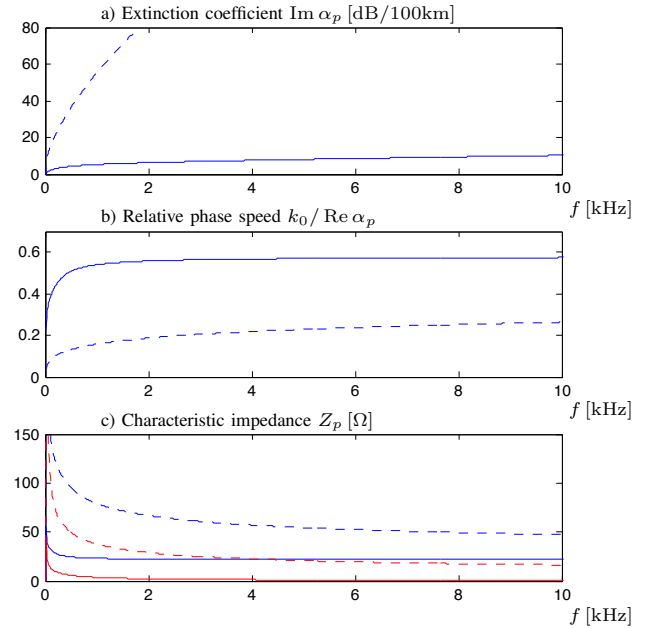
\begin{figure}[htb]
\begin{picture}(50,250)
\put(100,0){\makebox(50,230){\includegraphics[width=8cm]{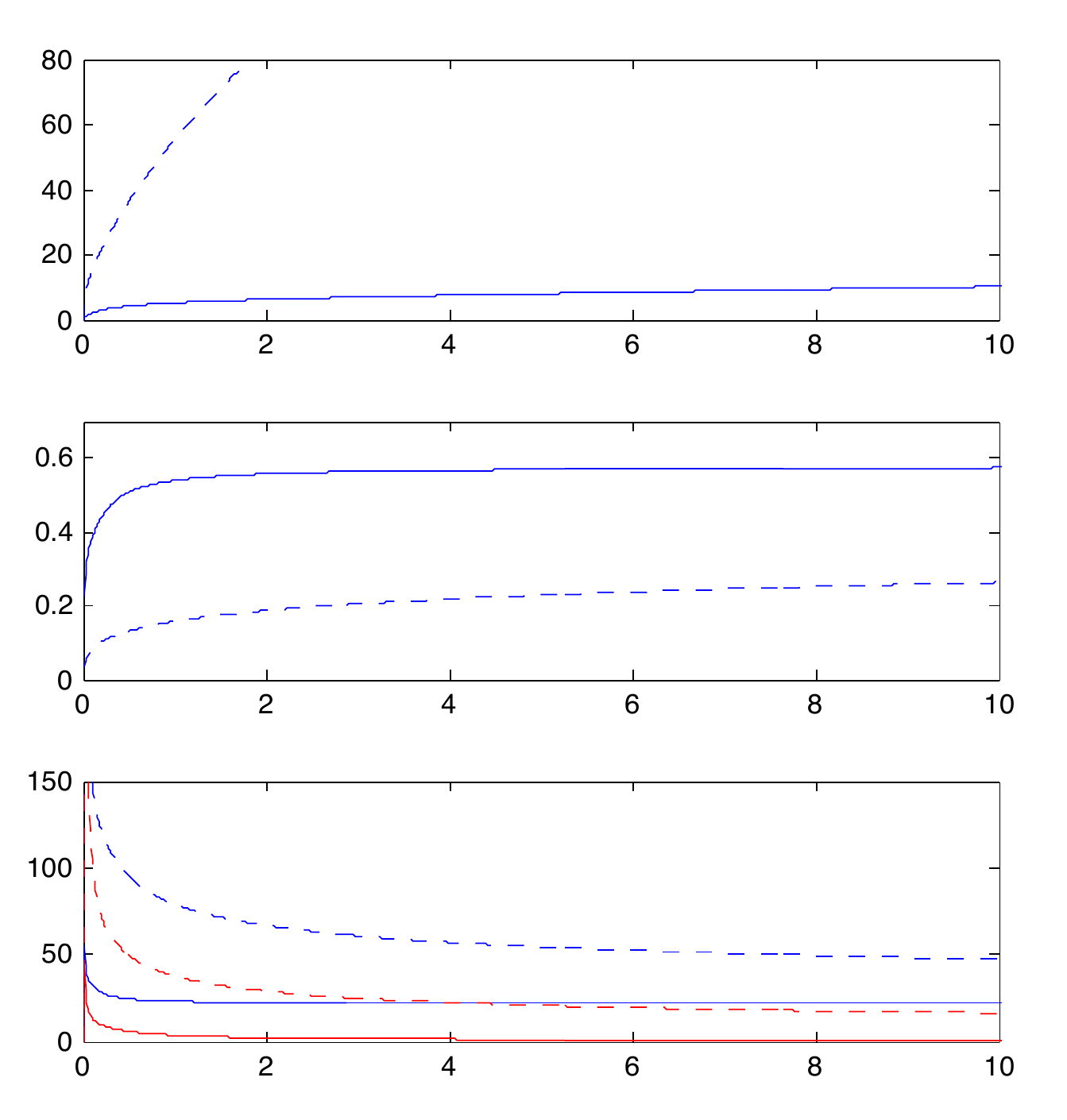}}} 
\put(50,225){\scriptsize a) Extinction coefficient  $\Im\alpha_p$\unit{[dB/100km]}}
\put(227,151){\scriptsize $f \unit{[kHz]}$}
\put(50,148){\scriptsize b) Relative phase speed $k_0/\Re\alpha_p$}
\put(227,75){\scriptsize $f \unit{[kHz]}$}
\put(50,71){\scriptsize c) Characteristic impedance $Z_p$\unit{[\ohm]}}
\put(227,-3){\scriptsize $f \unit{[kHz]}$}
\end{picture}
\caption{Illustration of wave propagation characteristics. The solid lines correspond to the $\textrm{TM}_{01}$ mode ($p=1$)
and the dashed lines to the $\textrm{TM}_{02}$ mode ($p=2$).
a) Extinction coefficient $\Im\alpha_p \cdot 2\cdot 10^{6}\log\eu \unit{[dB/100km]}$.
b) Relative phase speed $k_0/\Re\alpha_p$ (dimensionless).
c) Characteristic impedance $Z_p$\unit{[\ohm]}, where the blue lines show $\Re Z_p$ and the red lines $\Im Z_p$.}
\label{fig:matfig21}
\end{figure}

Once the pole $\alpha_p$ is found for a particular frequency $\omega$, 
the corresponding wave propagation characteristics of the cable can be studied in terms
of the extinction coefficient  $\Im\alpha_p$ and the relative phase speed $k_0/\Re\alpha_p$.
The characteristic impedance $Z_p$ can also be computed as outlined in section \ref{sect:charZ}.
In Fig.~\ref{fig:matfig21} is illustrated the wave propagation characteristics 
of the $\textrm{TM}_{01}$ and the $\textrm{TM}_{02}$ modes in the frequency range 0--10\unit{kHz}.

\begin{figure}[htb]
\begin{picture}(50,220)
\put(105,0){\makebox(50,200){\includegraphics[width=7cm]{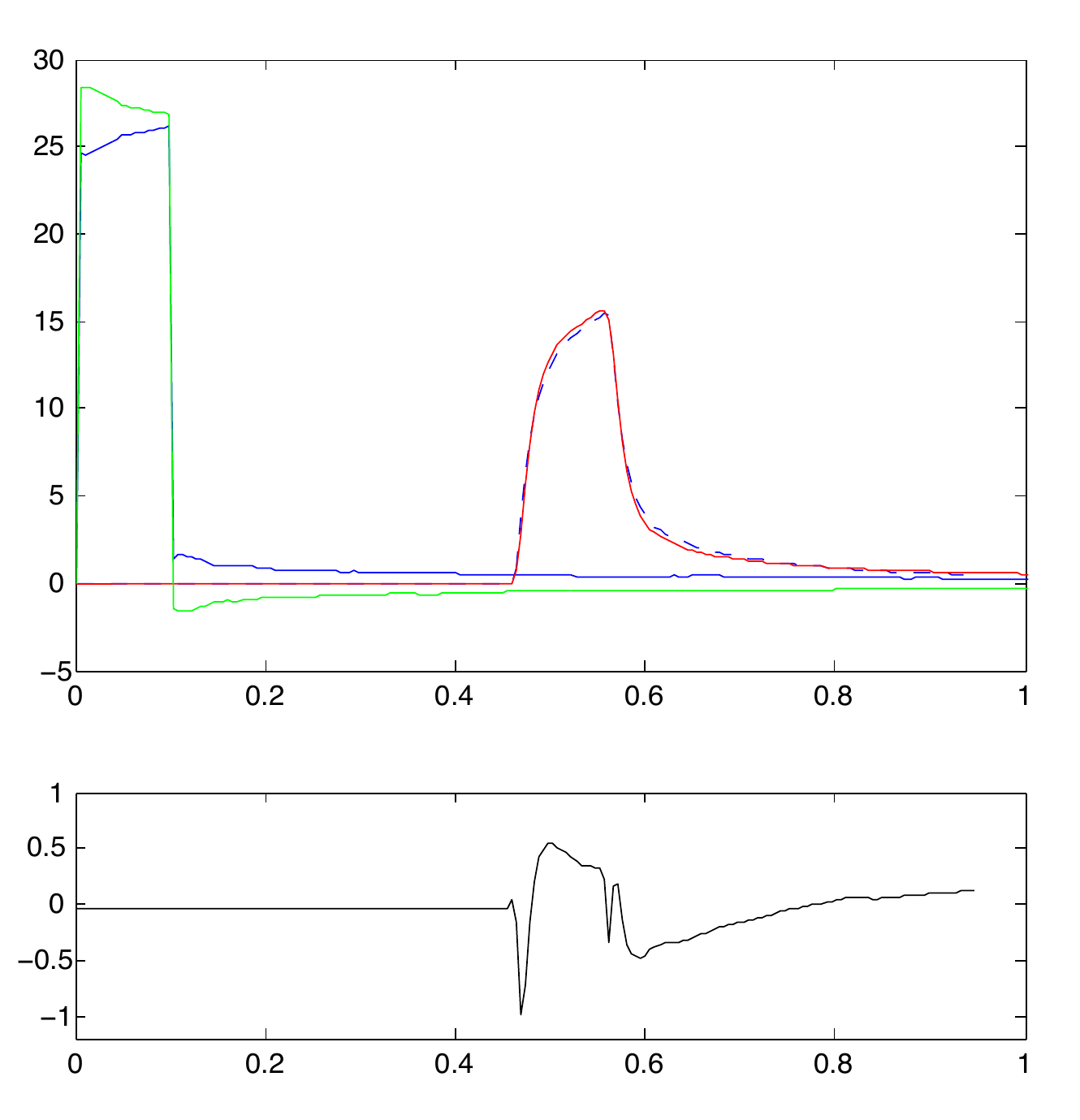}}} 
\put(75,196){\scriptsize a) Modeled and measured pulses\unit{[V]}}
\put(222,66){\scriptsize $t$ [\unit{ms}]}
\put(75,60){\scriptsize b) Remaining model error \unit{[V]}}
\put(222,-2){\scriptsize $t$ [\unit{ms}]}
\end{picture}
\caption{a) Modeling and measurements of pulse dispersion. Solid blue line: input voltage $v_{\rm in}(t)$.
Solid green line: input resistor voltage $Ri_{\rm in}(t)$. Dashed blue line: measured pulse $v^{\rm M}(t)$.
Solid red line: modeled pulse $v_1(t,z)$ at $z=81.8$\unit{km}. 
b) Remaining model error $v_1(t,z)-v^{\rm M}(t)$.}  
\label{fig:matfig31}
\end{figure}
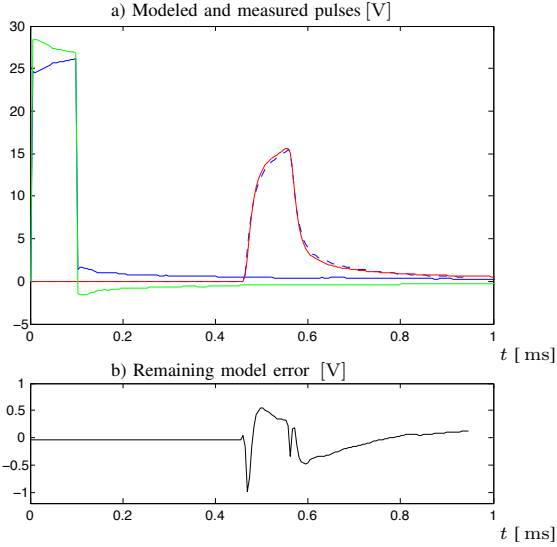

\begin{figure}[htb]
\begin{picture}(50,140)
\put(100,0){\makebox(50,120){\includegraphics[width=8cm]{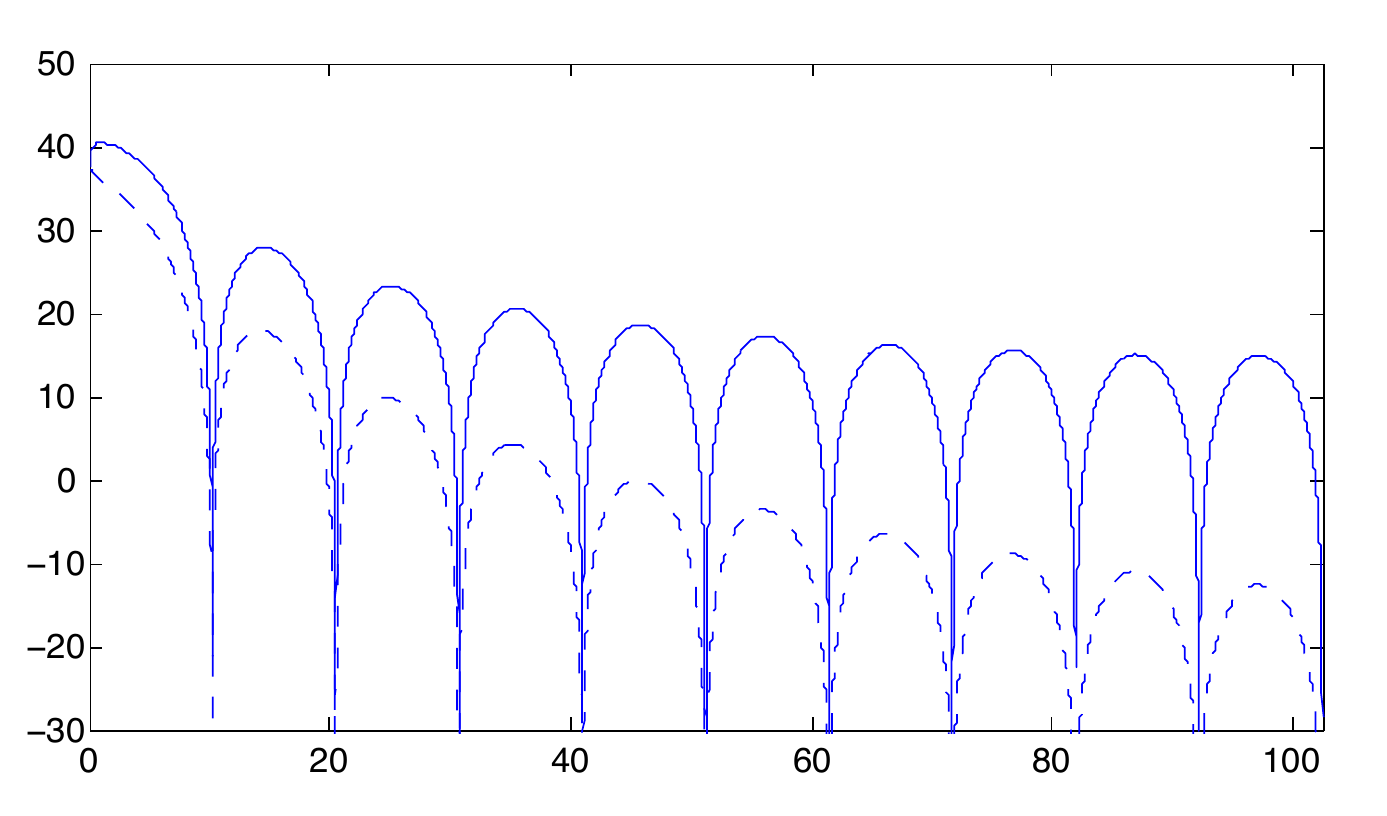}}} 
\put(50,120){\scriptsize Conductor current amplitude $|I_1|$\unit{[dB/1\mu As]}}
\put(227,-6){\scriptsize $f \unit{[kHz]}$}
\end{picture}
\caption{Conductor current based on calibrated magnetic frill generator excitation plotted for 0--100\unit{kHz}.
The plot shows the current amplitude of the $\textrm{TM}_{01}$ mode $I_1(f,z)$ at
$z=0$ (solid line) and at $z=81.8$\unit{km} (dashed line).}
\label{fig:matfig23}
\end{figure}

Once the contours ${\cal C}_p$ of the poles $\alpha_p$ are determined, the corresponding
modal contributions $I_p(\omega,z)$ of the conductor current can be computed numerically based 
on the integral \eqref{eq:intpdef}. The contribution from the branch-cut $I_{\rm br}(\omega,z)$ is similarly computed based
on the path of steepest descent for the integral in \eqref{eq:intbrdef2}. The excitation amplitude $M(\omega)$ is
calculated as in \eqref{eq:Mdef}, and where $I_{\rm in}(\omega)$ is the input current that has been measured at the cable for $z=0$.
The output voltage $V_1(\omega)$ of the cable corresponding to the dominating quasi-TEM $\textrm{TM}_{01}$ mode
is modeled by $V_1(\omega)=Z_1I_1(\omega,z)2R/(Z_1+R)$ where $R=25$\unit{\ohm} is the internal resistance of the measuring device,
and $z=81.8$\unit{km}. The frequency domain result is tapered with a suitable spectral window and then transformed to the time domain
by using the Inverse Fast Fourier Transform (IFFT) \cite{Oppenheim+Schafer1999}.
Here, all frequency domain modeling and measurements are executed with a 16384-point FFT and a 
Nyquist frequency of 102.4\unit{kHz}. The corresponding time domain measurements and modeling results are shown
in Fig.~\ref{fig:matfig31}. The spectral content of the current of the $\textrm{TM}_{01}$ mode  
$I_1(\omega,z)$ is shown in Fig.~\ref{fig:matfig23} for $z=0$ and $z=81.8$\unit{km}.

\begin{figure}[htb]
\begin{picture}(50,140)
\put(100,0){\makebox(50,120){\includegraphics[width=8cm]{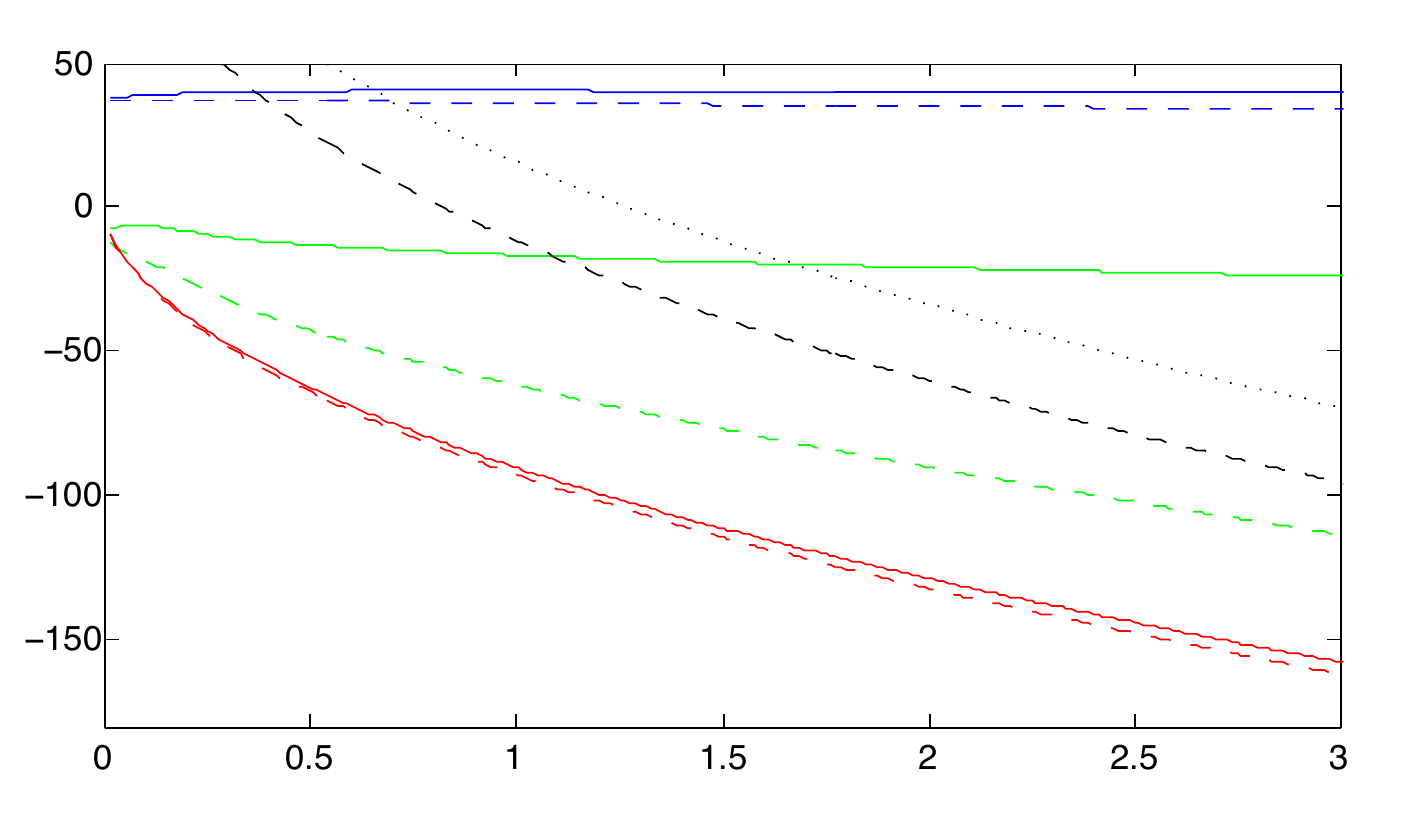}}} 
\put(30,120){\scriptsize Conductor current amplitudes $|I_1|$,  $|I_2|$,  $|I_{\rm br}|$,  $|I_{\rm br}^{\rm as}|$, $I_{\rm br}^{\rm ub}$\unit{[dB/1\mu As]}}
\put(227,-6){\scriptsize $f \unit{[kHz]}$}
\end{picture}
\caption{Comparison of conductor current contributions over the frequency range 0--3\unit{kHz}.
The plot shows the current amplitudes of the $\textrm{TM}_{01}$ mode $I_1(f,z)$ (blue lines), 
the $\textrm{TM}_{02}$ mode $I_2(f,z)$ (green lines) and the contribution from the branch-cut 
$I_{\rm br}(f,z)$ (red lines) at $z=0$ (solid lines) and at $z=81.8$\unit{km} (dashed lines), respectively.
The black dashed line shows the asymptotic approximation of the contribution from the branch-cut $I_{\rm br}^{\rm as}(f,z)$
and the black dotted line the corresponding upper bound $I_{\rm br}^{\rm ub}(f,z)$ at $z=81.8$\unit{km}.}
\label{fig:matfig24}
\end{figure}

\begin{figure}[htb]
\begin{picture}(50,140)
\put(100,0){\makebox(50,120){\includegraphics[width=8cm]{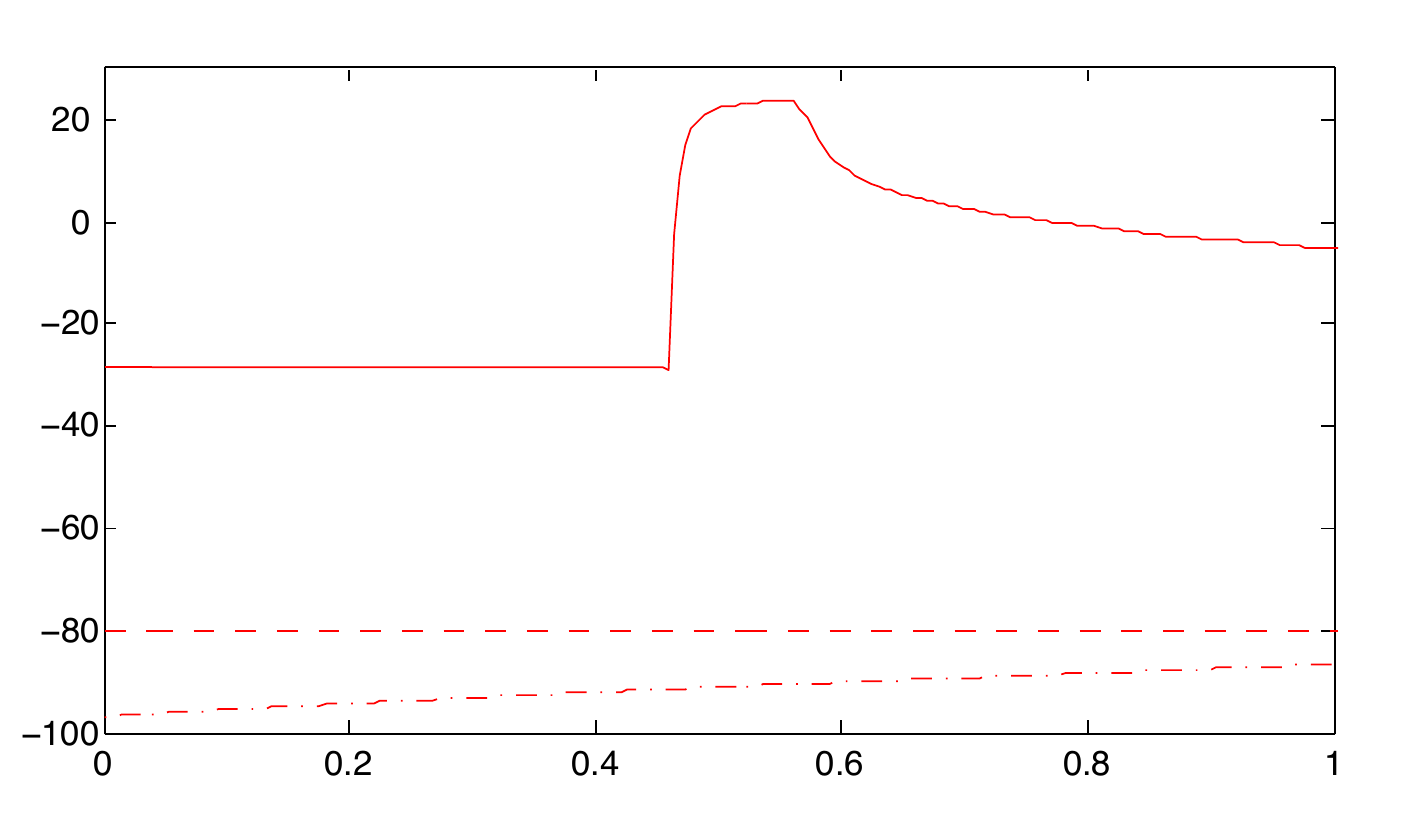}}} 
\put(45,119){\scriptsize Modeled pulse contributions $v_1$, $Ri_2$ and $Ri_{\rm br}$ \unit{[dB/1V]}}
\put(230,-6){\scriptsize $t$ [\unit{ms}]}
\end{picture}
\caption{Comparison of pulse contributions in the time domain. The solid line shows the $\textrm{TM}_{01}$ contribution $v_1(t,z)$,
the dashed line the $\textrm{TM}_{02}$ contribution $Ri_2(t,z)$ and the dash-dotted line the contribution from the branch-cut
$Ri_{\rm br}(t,z)$ for $z=81.8$\unit{km}, respectively.}
\label{fig:matfig36}
\end{figure}

Except for a better general understanding regarding the theory for non-discrete radiating modes of multilayered
open waveguide structures, the motivation for this work is also to study the potential contribution from higher order modes as well as from the branch-cut 
in a realistic scenario concerning the wave propagation on power cables. To this end, we argue that the simple magnetic frill generator
that is used here gives an excitation model that is sufficiently accurate to serve as a relevant example.
Hence, in Fig.~\ref{fig:matfig24} is shown a comparison of
the modal contributions $I_p(\omega,z)$ defined in \eqref{eq:intpdef} for $p=1,2$,
the contribution from the branch-cut $I_{\rm br}(\omega,z)$ defined in \eqref{eq:intbrdef2},
the asymptotic approximation $I_{\rm br}^{\rm as}(\omega,z)$ defined in \eqref{eq:intbrasdef}
and the corresponding upper bound $I_{\rm br}^{\rm ub}(\omega,z)$ defined in \eqref{eq:Ibrub}.
A numerical evaluation of the rest term $r(\alpha)$ has been carried out to determine the upper bound $M$
defined in \eqref{eq:restterm}. 
The results are shown for $z=0$ and $z=81.8$\unit{km}.
It is noted that the higher order modes and the contribution from the branch-cut
are only weakly excited, they contribute mainly at lower frequencies
and they are significantly damped in comparison to the dominant $\textrm{TM}_{01}$ mode.
Except for very low frequencies (less than 20\unit{Hz} or so), the $\textrm{TM}_{02}$ mode dominates
over the contribution from the branch-cut. 
It is also noted that since the branch-point is real valued (lossless exterior domain), 
the contribution from the branch-cut does not decay much with distance. 
In Fig.~\eqref{fig:matfig36} is shown the corresponding results in the time domain
plotted on a logarithmic (\unit{dB}) scale. It is observed that the pulse amplitude for higher order modes
as well as for the contribution from the branch-cut is more than 100\unit{dB} less than the contribution from
the dominant mode in the relevant timing interval.

\begin{figure}[htb]
\begin{picture}(50,140)
\put(100,0){\makebox(50,120){\includegraphics[width=8cm]{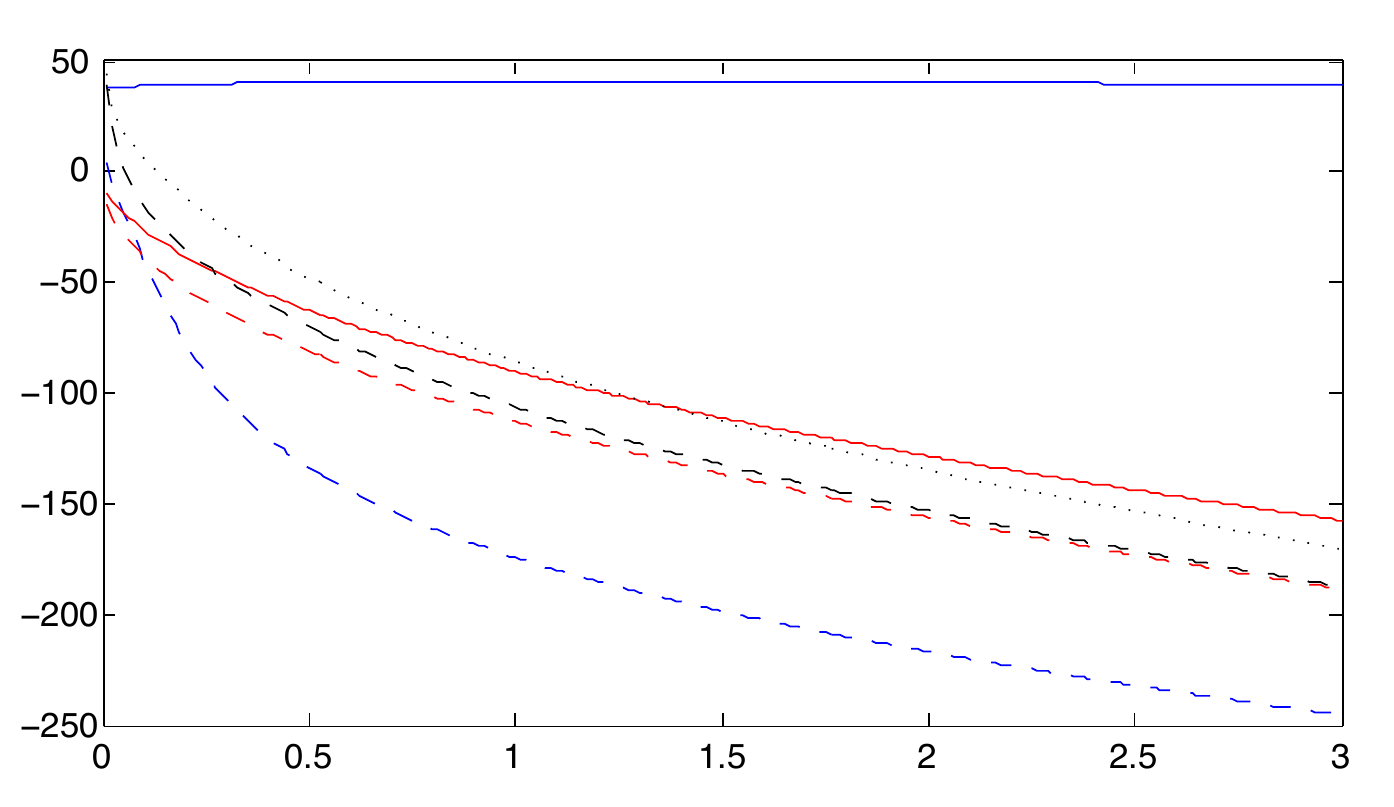}}} 
\put(30,120){\scriptsize Conductor current amplitudes $|I_1|$,  $|I_{\rm br}|$,  $|I_{\rm br}^{\rm as}|$, $I_{\rm br}^{\rm ub}$\unit{[dB/1\mu As]}}
\put(228,-6){\scriptsize $f \unit{[kHz]}$}
\end{picture}
\caption{Comparison of conductor current contributions over the frequency range 0--3\unit{kHz}.
The plot shows the current amplitudes of the $\textrm{TM}_{01}$ mode $I_1(f,z)$ (blue lines) and the contribution from the branch-cut
$I_{\rm br}(f,z)$ (red lines) at $z=0$ (solid lines) and at $z=50\cdot 81.8$\unit{km} (dashed lines), respectively.
The black dashed line shows the asymptotic approximation of the contribution from the branch-cut $I_{\rm br}^{\rm as}(f,z)$
and the black dotted line the corresponding upper bound $I_{\rm br}^{\rm ub}(f,z)$ at $z=50\cdot 81.8$\unit{km}.}
\label{fig:matfig25}
\end{figure}

It is also interesting to compare the various modal constituents 
at extremely long distances. In Fig.~\ref{fig:matfig25} is shown the corresponding results
regarding the $\textrm{TM}_{01}$ mode $I_1(\omega,z)$ and the contribution from the branch-cut
$I_{\rm br}(\omega,z)$ at $z=0$ and at $z=50\cdot 81.8$\unit{km}. The situation is now quite different and
for frequencies above some 200\unit{Hz} the contribution
from the branch-cut dominates over the $\textrm{TM}_{01}$ mode.
It is also at these extremely long distances that the asymptotic approximation
$I_{\rm br}^{\rm as}(\omega,z)$ becomes very close to the numerically computed integral $I_{\rm br}(\omega,z)$.


\section{Summary}\label{sect:Summary}

A detailed modeling and analysis have been presented regarding the dispersion characteristics of multilayered open coaxial waveguides.
The main application is with transient signal analysis for very long power cables.
An electromagnetic model has been developed which is based on a layer recursive computation of determinants in connection
with a magnetic frill generator excitation.
The layer recursive formulation enables a stable and efficient numerical computation of the related dispersion functions as well as
a detailed analysis regarding the analytic and asymptotic properties of the associated determinants.
Modal contributions as well as the contribution from the associated branch-cut (non-discrete radiating modes) 
have been defined and analyzed.
A concrete example have been included based on measurements and modeling of pulse propagation on
an 82 km long HVDC power cable.
In this example, it is concluded that the contribution from the second TM mode as well as from 
the branch-cut is negligible for all practical purposes. However, it has also been shown that for extremely long power cables the
contribution from the branch-cut can in fact dominate over the quasi-TEM mode for some frequency intervals.

\appendices
\section{Definition of cylindrical vector waves}\label{app:cyldef}
The Fourier transform of the electric field over the longitudinal space coordinate $z$ is defined here by 
\begin{equation}\label{eq:Ftransfalpha}
\left\{\begin{array}{l}
\vec{E}(\vec{\rho},\alpha)=\displaystyle\int_{-\infty}^{\infty}\vec{E}(\vec{r})\eu^{-\iu\alpha z}\diff z,\vspace{0.2cm} \\
\vec{E}(\vec{r})=\displaystyle\frac{1}{2\pi}\int_{-\infty}^{\infty}\vec{E}(\vec{\rho},\alpha)\eu^{\iu\alpha z}\diff \alpha,
\end{array}\right.
\end{equation}
and similarly for the magnetic field. It is assumed that all fields are analytic functions of $\alpha=\alpha^\prime+\iu\alpha^{\prime\prime}$
for $\alpha^{\prime\prime}_-<\Im\alpha<\alpha^{\prime\prime}_+$ where $\alpha^{\prime\prime}_-<0$ and $\alpha^{\prime\prime}_+>0$.

In the cylindrical coordinate system the fields are periodic in the angular coordinate $\phi$.
The corresponding Fourier series expansion of the electric field is defined here by
\begin{equation}\label{eq:Fser}
\left\{\begin{array}{l}
\vec{E}_m(\vec{\rho},\alpha)=\displaystyle\frac{1}{2\pi}\int_{-\pi}^{\pi}\vec{E}(\vec{\rho},\alpha)\eu^{-\iu m\phi}\diff \phi, \vspace{0.2cm} \\
\vec{E}(\vec{\rho},\alpha)=\displaystyle\sum_{m=-\infty}^{\infty}\vec{E}_m(\vec{\rho},\alpha)\eu^{\iu m\phi},
\end{array}\right.
\end{equation}
and similarly for the magnetic field. 
Note that the Fourier integral in (\ref{eq:Fser}) is defined to operate only on the cylindrical coordinates of the vector
field, and does not operate on the space dependent unit vectors $\hat{\vec{\rho}}$ and $\hat{\vec{\phi}}$.

Consider a homogeneous and isotropic cylindrical region with relative permittivity $\epsilon$, relative permeability $\mu$ and
wavenumber $k=k_0\sqrt{\mu\epsilon}$.
The cylindrical vector waves are defined here by
\begin{equation}\label{eq:cylvecdef}
\left\{\begin{array}{l}
\vec{\chi}_{1m}(\vec{r},\alpha)=\displaystyle\frac{1}{\kappa}\nabla\times\left(\hat{\vec{z}}\psi_m(\kappa\rho)\eu^{\iu m\phi}\eu^{\iu\alpha z}\right), \vspace{0.2cm} \\
\vec{\chi}_{2m}(\vec{r},\alpha)=\displaystyle\frac{1}{k}\nabla\times\vec{\chi}_{1m}(\vec{r},\alpha),
\end{array}\right.
\end{equation}
where $\psi_m(\kappa\rho)$ is a regular Bessel function or a Hankel function of the first kind, both of 
order $m$, see also \cite{Bostrom+Kristensson+Strom1991,Collin1991}.
Here, $\alpha$ is the longitudinal wavenumber and $\kappa=\sqrt{k^2-\alpha^2}$ the transverse wavenumber where
the square root is chosen such that $0<\arg\kappa\leq \pi$ and hence $\Im \kappa\geq 0$. 
The following notation will be used
\begin{equation}\label{eq:cylvecexpr2}
\vec{\chi}_{\tau m}(\vec{r},\alpha)=\vec{\chi}_{\tau m}(\vec{\rho},\alpha)\eu^{\iu m\phi}\eu^{\iu\alpha z},
\end{equation}
where $\tau=1,2$, and the vectors $\vec{\chi}_{\tau m}(\vec{\rho},\alpha)$ are given explicitly 
in cylindrical coordinates as
\begin{equation}\label{eq:cylvecexpr}
\left\{\begin{array}{l}
\vec{\chi}_{1m}(\vec{\rho},\alpha)=\displaystyle \hat{\vec{\rho}}\frac{\iu m}{\kappa\rho}\psi_{m}(\kappa\rho)
-\hat{\vec{\phi}}\psi_{m}^{\prime}(\kappa\rho),  \vspace{0.2cm} \\
\vec{\chi}_{2m}(\vec{\rho},\alpha)=\displaystyle \hat{\vec{\rho}}\frac{\iu\alpha}{k}\psi_{m}^{\prime}(\kappa\rho)
-\hat{\vec{\phi}}\frac{m\alpha}{k\kappa\rho}\psi_{m}(\kappa\rho)
+\hat{\vec{z}}\frac{\kappa}{k}\psi_{m}(\kappa\rho),
\end{array}\right.
\end{equation}
and where $\psi^\prime(\cdot)$ denotes a differentiation with respect to the argument. 

It can be shown by direct calculation that $\nabla\times \vec{\chi}_{2m}(\vec{r},\alpha)=k\vec{\chi}_{1m}(\vec{r},\alpha)$.
The following curl properties are thus obtained
\begin{equation}\label{eq:chi12cross}
\nabla\times \vec{\chi}_{\tau m}(\vec{r},\alpha)=k\vec{\chi}_{\bar{\tau} m}(\vec{r},\alpha), 
\end{equation}
for  $\tau=1,2$, and where $\bar{\tau}$ denotes the complement of $\tau$ ($\bar{1}=2$ and $\bar{2}=1$).
It follows from \eqref{eq:chi12cross} that the second order curl properties are given by
\begin{equation}\label{eq:chi12cross2}
\nabla\times\nabla\times \vec{\chi}_{\tau m}(\vec{r},\alpha)=k^2\vec{\chi}_{\tau m}(\vec{r},\alpha),
\end{equation}
for  $\tau=1,2$.
It follows also from (\ref{eq:chi12cross}) that the two cylindrical vector waves are solenoidal with 
$\nabla\cdot \vec{\chi}_{\tau m}(\vec{r},\alpha)=0$,
and from (\ref{eq:chi12cross2}) that they satisfy the vector Helmholtz wave equation
\begin{equation}\label{eq:Maxwellcylvec2}
\nabla^2\vec{\chi}_{\tau m}(\vec{r},\alpha)+k^2\vec{\chi}_{\tau m}(\vec{r},\alpha)=\vec{0}, 
\end{equation}
for  $\tau=1,2$. 

Let the regular and the outgoing (radiating) cylindrical vector waves $\vec{v}_{\tau m}(\vec{r},\alpha)$ 
and $\vec{u}_{\tau m}(\vec{r},\alpha)$ be defined as in (\ref{eq:cylvecdef}) by using the regular Bessel functions and 
the Hankel functions of the first kind, ${\rm J}_m(\kappa\rho)$ and ${\rm H}_m^{(1)}(\kappa\rho)$, 
respectively.
The electric and magnetic fields can then generally be expanded as
\begin{multline}\label{eq:cylvecmodelE}
\vec{E}(\vec{r})=\displaystyle\frac{1}{2\pi}\int_{-\infty}^{\infty}\sum_{m=-\infty}^{\infty}\vec{E}_m(\vec{\rho},\alpha)\eu^{\iu m\phi}\eu^{\iu\alpha z}\diff\alpha \\
=\displaystyle\frac{1}{2\pi}\int_{-\infty}^{\infty}\sum_{m=-\infty}^{\infty}\sum_{\tau=1}^{2}\left[
a_{\tau m}(\alpha)\vec{v}_{\tau m}(\vec{r},\alpha) \right. \\
\left. +b_{\tau m}(\alpha)\vec{u}_{\tau m}(\vec{r},\alpha) \right]\diff\alpha,
\end{multline}
and
\begin{multline}\label{eq:cylvecmodelH}
\vec{H}(\vec{r})=\displaystyle\frac{1}{2\pi}\int_{-\infty}^{\infty}\sum_{m=-\infty}^{\infty}\vec{H}_m(\vec{\rho},\alpha)\eu^{\iu m\phi}\eu^{\iu\alpha z}\diff\alpha \\
=\displaystyle \frac{1}{\iu\eta_0\eta} \frac{1}{2\pi}\int_{-\infty}^{\infty}\sum_{m=-\infty}^{\infty}\sum_{\tau=1}^{2}\left[
a_{\tau m}(\alpha)\vec{v}_{\bar{\tau} m}(\vec{r},\alpha)\right. \\
\left. +b_{\tau m}(\alpha)\vec{u}_{\bar{\tau} m}(\vec{r},\alpha)\right]\diff\alpha,
\end{multline}
where $\vec{H}=\frac{1}{\iu k_0\eta_0\mu}\nabla\times\vec{E}$ and (\ref{eq:chi12cross}) have been used, 
see also \cite{Bostrom+Kristensson+Strom1991,Collin1991}. 
Here, $a_{\tau m}(\alpha)$ and $b_{\tau m}(\alpha)$ are complex valued expansion coefficients 
with the same dimension as the electric field (\unit{Vs/m}), and which can be determined by applying the appropriate boundary conditions.
The expansions (\ref{eq:cylvecmodelE}) and (\ref{eq:cylvecmodelH}) are valid in homogeneous and source-free cylindrical regions or layers. 
For the interior region containing the origin ($\rho=0$), the expansions
(\ref{eq:cylvecmodelE}) and (\ref{eq:cylvecmodelH}) consist only of regular vector waves $\vec{v}_{\tau m}(\vec{r})$.
For the exterior region (containing $\rho=\infty$) the expansions (\ref{eq:cylvecmodelE}) and (\ref{eq:cylvecmodelH}) 
consist only of outgoing (radiating) vector waves $\vec{u}_{\tau m}(\vec{r})$.

Field components with coefficients $a_{1m}$ and $b_{1m}$ ($\tau=1$) are referred to as Transverse Electric (TE), and
fields with coefficients $a_{2m}$ and $b_{2m}$ ($\tau=2$) are referred to as Transverse Magnetic (TM). 
Finally, it is noted that the transformed field can be written as
\begin{multline}
\vec{E}_m(\vec{\rho},\alpha)=\sum_{\tau=1}^{2}a_{\tau m}(\alpha)\vec{v}_{\tau m}(\vec{\rho},\alpha)+b_{\tau m}(\alpha)\vec{u}_{\tau m}(\vec{\rho},\alpha) \\
=\hat{\vec{\rho}}E_{m\rho}(\rho,\alpha)+\hat{\vec{\phi}}E_{m\phi}(\rho,\alpha)+\hat{\vec{z}}E_{mz}(\rho,\alpha),
\end{multline}
where $E_{m\rho}(\rho,\alpha)$, $E_{m\phi}(\rho,\alpha)$ and $E_{mz}(\rho,\alpha)$ denote the cylindrical coordinates
of $\vec{E}_m(\vec{\rho},\alpha)$, and similarly for the magnetic field.

\section{Recursive computation of determinants}\label{sect:recdet}
The determinants $\det{\bf A}(\omega,\alpha)$ and $\det{\bf B}(\omega,\alpha)$ related to the linear system of equations
in \eqref{eq:Matconds1} through \eqref{eq:Matconds3} have the general structure 
\begin{equation}{\scriptsize 
\left|\begin{array}{cccccccc}
-x_1^1  & x_1^3  & x_1^4 &   &   &   &   &  \vspace{0.2cm} \\
-y_1^1  & y_1^3  & y_1^4 &   &   &   &   &  \vspace{0.2cm} \\
  & -x_2^1  & -x_2^2 & x_2^3  & x_2^4 &   &   &       \vspace{0.2cm} \\
  & -y_2^1  & -y_2^2 & y_2^3  & y_2^4 &   &   &       \vspace{0.2cm} \\
  &   &   & \ddots  &  \ddots &   &   &         \vspace{0.2cm} \\
  &  &  & -x_{N-1}^1  & -x_{N-1}^2 & x_{N-1}^3  & x_{N-1}^4 &         \vspace{0.2cm} \\
  &  &  & -y_{N-1}^1  & -y_{N-1}^2 & y_{N-1}^3  & y_{N-1}^4 &        \vspace{0.2cm} \\
  &  &  &  &  & -x_{N}^1  & -x_{N}^2  & x_{N}^4          \vspace{0.2cm} \\
  &  &  &  &  & -y_{N}^1  & -y_{N}^2  & y_{N}^4         
\end{array}\right|}
\end{equation}
where the elements of ${\bf A}(\omega,\alpha)$ are given by
\begin{equation}\label{eq:x1234def}
\left\{\begin{array}{l}
x_i^1=\kappa_i{\rm J}_0(\kappa_i\rho_i), \quad i=1,\ldots,N, \vspace{0.2cm} \\
x_i^2=\kappa_i{\rm H}_0^{(1)}(\kappa_i\rho_i), \quad i=2,\ldots,N, \vspace{0.2cm} \\
x_i^3=\kappa_{i+1}{\rm J}_0(\kappa_{i+1}\rho_i), \quad i=1,\ldots,N-1, \vspace{0.2cm} \\
x_i^4=\kappa_{i+1}{\rm H}_0^{(1)}(\kappa_{i+1}\rho_i), \quad i=1,\ldots,N, 
\end{array}\right.
\end{equation}
and
\begin{equation}\label{eq:y1234def}
\left\{\begin{array}{l}
y_i^1=\epsilon_i{\rm J}_1(\kappa_i\rho_i), \quad i=1,\ldots,N, \vspace{0.2cm} \\
y_i^2=\epsilon_i{\rm H}_1^{(1)}(\kappa_i\rho_i), \quad i=2,\ldots,N, \vspace{0.2cm} \\
y_i^3=\epsilon_{i+1}{\rm J}_1(\kappa_{i+1}\rho_i), \quad i=1,\ldots,N-1, \vspace{0.2cm} \\
y_i^4=\epsilon_{i+1}{\rm H}_1^{(1)}(\kappa_{i+1}\rho_i), \quad i=1,\ldots,N, 
\end{array}\right.
\end{equation}
and similarly for the matrix ${\bf B}(\omega,\alpha)$ where $-x_1^1=1$ and $y_1^1=0$. Straightforward modifications can be
implemented to adapt for the case with exponential scalings as described in section \ref{sect:expscale}.

Let $f_i$ and $g_i$ denote auxiliary determinants for $i=1,\ldots,N$.
By defining the first order determinants as
\begin{equation}
\left\{\begin{array}{l}
f_1=-x_1^1, \vspace{0.2cm} \\
g_1=-y_1^1,
\end{array}\right.
\end{equation}
the second order determinants can be computed as
\begin{multline}
f_2=\left|\begin{array}{ccc}
f_1  & x_1^3  & x_1^4  \vspace{0.2cm} \\
g_1  &  y_1^3  & y_1^4  \vspace{0.2cm} \\
0   &  -x_2^1 & -x_2^2
\end{array}\right|  \\
=(x_2^1y_1^4-x_2^2y_1^3)f_1+(x_2^2x_1^3-x_2^1x_1^4)g_1,
\end{multline}
and
\begin{multline}
g_2=\left|\begin{array}{ccc}
f_1  & x_1^3  & x_1^4  \vspace{0.2cm} \\
g_1  &  y_1^3  & y_1^4  \vspace{0.2cm} \\
0   &  -y_2^1 & -y_2^2
\end{array}\right|  \\
=(y_2^1y_1^4-y_2^2y_1^3)f_1+(y_2^2x_1^3-y_2^1x_1^4)g_1.
\end{multline}
Similarly, it can be shown that the $i$th order determinants can be computed recursively as
\begin{multline}
f_i=\left|\begin{array}{ccccc}
\ddots &   &    &    & \vspace{0.2cm} \\
          & -x_{i-1}^1 & -x_{i-1}^2 & x_{i-1}^3  & x_{i-1}^4 \vspace{0.2cm} \\
         & -y_{i-1}^1 & -y_{i-1}^2 & y_{i-1}^3  & y_{i-1}^4 \vspace{0.2cm} \\
         &   &    &  -x_i^1 &   -x_i^2
\end{array}\right|  \\
=(x_i^1y_{i-1}^4-x_i^2y_{i-1}^3)f_{i-1}+(x_i^2x_{i-1}^3-x_i^1x_{i-1}^4)g_{i-1},
\end{multline}
and
\begin{multline}
g_i=\left|\begin{array}{ccccc}
\ddots &   &    &    & \vspace{0.2cm} \\
          & -x_{i-1}^1 & -x_{i-1}^2 & x_{i-1}^3  & x_{i-1}^4 \vspace{0.2cm} \\
         & -y_{i-1}^1 & -y_{i-1}^2 & y_{i-1}^3  & y_{i-1}^4 \vspace{0.2cm} \\
         &   &    &  -y_i^1 &   -y_i^2
\end{array}\right|  \\
=(y_i^1y_{i-1}^4-y_i^2y_{i-1}^3)f_{i-1}+(y_i^2x_{i-1}^3-y_i^1x_{i-1}^4)g_{i-1}.
\end{multline}
The final determinant $\det{\bf X}$ ($\det{\bf A}(\omega,\alpha)$ or $\det{\bf B}(\omega,\alpha)$)
is computed as
\begin{equation}\label{eq:finaldet}
\det{\bf X}=y_N^4f_N-x_N^4g_N,
\end{equation}
where
\begin{equation}
\left\{\begin{array}{l}\label{eq:recdef}
f_i=A_if_{i-1}+B_ig_{i-1},  \vspace{0.2cm} \\
g_i=C_if_{i-1}+D_ig_{i-1},
\end{array}\right.
\end{equation}
and
\begin{equation}\label{eq:ABCDdef}
\left\{\begin{array}{l}
A_i= x_i^1y_{i-1}^4-x_i^2y_{i-1}^3, \vspace{0.2cm} \\
B_i=x_i^2x_{i-1}^3-x_i^1x_{i-1}^4,  \vspace{0.2cm} \\
C_i=y_i^1y_{i-1}^4-y_i^2y_{i-1}^3,  \vspace{0.2cm} \\
D_i=y_i^2x_{i-1}^3-y_i^1x_{i-1}^4,
\end{array}\right.
\end{equation}
for $i=2,\ldots,N$.

\bibliographystyle{teorel}


\begin{IEEEbiographynophoto}{Sven~Nordebo}
received the M.S.\ degree in electrical engineering from the Royal Institute of Technology, Stockholm, Sweden, in 1989, 
and the Ph.D.\ degree in signal processing from Lule{\aa} University of Technology, Lule{\aa}, Sweden, in 1995. 
Since 2002 he is a Professor of Signal Processing at the 
School of Computer Science, Physics and Mathematics, Linn\ae us University. 
His research interests are in statistical signal processing, electromagnetic wave propagation, inverse problems and imaging.
\end{IEEEbiographynophoto}\vspace{-0.5cm}

\begin{IEEEbiographynophoto}{G\"{o}khan~Cinar}
G\"{o}khan Cinar received the B.S. and M.S. degrees in electronics and
telecommunication engineering from Istanbul Technical University, Turkey, in
1998 and 2001, respectively, and the Ph.D. degree in electronics engineering
from Gebze Institute of Technology, Turkey, in 2004. He received his docent
degree in 2010 and since 2011 he is an Associate Professor of
Electromagnetics at the Electronics Engineering Department, Gebze Institute
of Technology. His research interests are in scattering and propagation of
electromagnetic and acoustic waves.
\end{IEEEbiographynophoto}\vspace{-0.5cm}

\begin{IEEEbiographynophoto}{Stefan~Gustafsson}
received the M.S. degree in mathematics from the Linn\ae us University,
V\"{a}xj\"{o}, Sweden, in 2005. Between 2007 and 2010 he worked as a test engineer in a high voltage
laboratory at ABB High Voltage Cables in Karlskrona, Sweden. He is now a Ph.D. student at
the Linn\ae us University.
\end{IEEEbiographynophoto}\vspace{-0.5cm}

\begin{IEEEbiographynophoto}{B\"{o}rje~Nilsson}
received the M.S. degree in mathematics and physics and the Ph.D. degree in theoretical physics from G\"{o}teborg University, 
Sweden, in 1971 and 1980, respectively. Since 2005 he is a Professor of Mathematical Physics at the School of Computer Science, 
Physics and Mathematics at Linn\ae us University. His main research interest is mathematical modelling of wave phenomena.
\end{IEEEbiographynophoto}\vspace{-0.5cm}

\end{document}